# The Human Capital Accumulation at Research Infrastructures: Reexamining Wage Returns to Training, Models, Interpretation, and Magnitude[1]

Erica Delugas[a], Francesco Giffoni[a], Emanuela Sirtori[a], Johannes Gutleber[b]

Affiliations:

[a] CSIL, Corso Monforte 15, 20122, Milan, Italy

[b]CERN, European Organization for Nuclear Research, Esplanade des Particules 1, 1211 Geneva 23, Switzerland

---

## Abstract

We design a research strategy to measure wage returns of research training targeting early-career researchers (ECRs) at research infrastructures (RIs). Grounded in established economic models of education and training, our strategy improves upon existing studies on the same topic in labour market relevance, scope, and economic modelling. We draw on a survey of ECRs at the European Organisation for Nuclear Research (CERN) and find that CERN research training increases ECRs' wages by 7% on average, ranging from 2 to 10%. Wage gains materialise early in their careers, typically within the first decade of employment. Wage returns are driven by hard skills – such as software development, data analysis, and problem-solving capacity – as well as sought-after soft skills, such as communication, leadership, and networking. Our findings suggest that ECRs' wage returns primarily reflect productivity improvements rather than the signalling effect of CERN affiliation. Our research applies to other RIs. We conclude by discussing methodological considerations, policy implications, and avenues for future research.

Keywords: wage returns to training, research infrastructures, early-career researchers, human capital accumulation, skills acquisition

---

*Corresponding author:

Email addresses: delugas@csilmilano.com* (Erica Delugas), giffoni@csilmilano.com (Francesco Giffoni), sirtori@csilmilano.com (Emanuela Sirtori), johannes.gutleber@cern.ch (Johannes Gutleber)

---

[1]The title of this paper borrows from the article's title "Reexamining the returns to training: Functional form, magnitude, and interpretation" by Frazis and Loewenstein (2005).

**Funding and Disclaimer**: The research presented in this paper was funded by the Future Circular Collider Innovation Study (FCCIS), a project supported by the European Union's Horizon 2020 research and innovation programme under grant No. 951754. The content of this paper reflects the views of the authors only, and neither CERN nor the European Commission can be held responsible for any use that may be made of the information contained herein.

**Acknowledgments.** We wish to thank Christian Caron (Springer Nature) and Armin Ilg (CERN) for their help to enrich and strengthen the content of this paper.

# 1. Introduction

The human capital theory states that the innate endowment of individual skills can be expanded during the life cycle through education and training, which are considered investments that will pay off in terms of productivity and earnings (Deming and Silliman, 2024; Demig 2022; Holmes, 2017; Grosemans et al., 2017; Frazis and Loewenstein, 2015). Today, research infrastructures (hereafter RIs) across various scientific fields – such as large particle accelerators, synchrotrons, outer space probes, cutting-edge technology centres, and cross-country data clouds – serve as critical loci where human and social capital are most effectively accelerated through the production of cutting-edge research, even more than universities (ESFRI, 2021; Catalano et al., 2021; OECD, 2019; Florio 2019).[2] In a world where economies are becoming increasingly complex and pressing issues such as climate change, health, energy, and sustainable development are global in nature, RIs ensure scientific excellence through the collaborative efforts of scientists from diverse universities, research centres, countries, fields, and backgrounds. This enables discoveries, advances in knowledge, and technological development (Gutleber and Charitos, 2025; Draghi 2024b; Castelnovo et al. 2024; Rossi et al., 2020).

The interest in training and job collaborations with RIs has grown over time as they offer a unique opportunity to enhance skills, build a reputation (Catalano et al., 2021), especially for early-career researchers (hereafter ECRs),[3] and expand both the demand for and professional networks of researchers in regions by attracting talents (Draghi, 2024a; Letta, 2024). In this paper, we address three main questions: (1) What are the wage returns for ECRs who spend a period conducting research at RIs? (2) Are the wage returns driven by the skills acquired during that training period? If so, which skills? (3) Does the reputational effect, rather than productivity gains, play a role? We answer these questions by analysing research training opportunities at the European Organisation for Nuclear Research (CERN) in Geneva, Switzerland, as a case study. It is an ideal case due to its cutting-edge research, structured programmes, global networking (including industry partnerships), and the development of transferable skills applicable in other sectors, such as data science, computational skills, project management, and team leadership (Anderson et al., 2013; Camporesi, 2001). Since 1993, nearly 37,000 ECRs have participated in experiments at CERN, including those within the Large Hadron Collider scientific programme (Catalano et al., 2018). By studying CERN, this paper aims to explore how world-class research training influences professional trajectories, skill development, and salary outcomes, offering valuable lessons for other RIs and policymakers (see, for instance, Ecchia et al., 2021).

[2]In this paper, we adopt the European Strategy Forum on Research Infrastructure (ESFRI) definition of RIs. They are "facilities, resources and related services that are used by the scientific community to conduct top-level research in their respective fields and covers major scientific equipment; knowledge-based resources such as collections, archives or structures for scientific information; enabling Information and Communications Technology-based infrastructures such as Grid, computing, software and communication, or any other entity of a unique nature essential to achieve excellence in research. Such infrastructures may be 'single-sited' or 'distributed'. See https://ec.europa.eu/info/research-and-innovation/strategy/european-research-infrastructures_en.

[3] According to Panel EFCA (2022), an early-career researcher is defined as PhD students and postdocs, either with a non-permanent contract or with up to 8 years after obtaining the PhD.

On top of that, we discuss estimation models, interpretation, and magnitude of wage returns to training vis-à-vis the benchmark human capital literature linking earning, training and work experience (e.g., Adda and Dustmann, 2023; Frazis and Loewenstein, 2015; 2005; Leuven and Oosterbeek, 2004; Acemoglu and Pischke, 1999) and the state-of-the-art in the literature of RIs on this topic (Ecchia et al., 2021; Catalano et al., 2021; Florio, 2019; Camporesi et al., 2017; Battistoni et al., 2016; Florio et al, 2016; Anderson et al., 2013).

We attain three main results:

1. We estimated that training in a research project at CERN yields a significant 7% increase in ECR's yearly wages in the baseline estimation, which aligns with the literature on wage returns to training (see Section 2.2).
2. The acquisition or the improvement of skills during the activities at CERN drives the post-training wage returns once in work. We find that hard skills, such as technical competencies (e.g., software development and data analysis), scientific expertise, and problem-solving capacity, are key determinants of higher wages. Additionally, soft skills, including networking, communication, and leadership, play an important role.
3. Wage returns arise because new or improved skills enhance the ECRs' existing human capital, making them more productive in the workplace due to the additional capabilities generated. As a result, we argue that productivity-related effects outweigh the prestige and reputational mechanisms that CERN affiliation might signal in the labour market.

Our contribution to the literature is manifold. First, it is methodological. To the best of our knowledge, this is the first study attempting to estimate the ECRs' wage returns to training at RIs employing the methods of the applied microeconomic literature of the human capital theory, such as Mincerian-like earning functions (Chiswick, 2024; Holmes, 2017; Mincer 1974). In this perspective, we bridge the novel stream of research on the socioeconomic impact of contemporary big-science projects[4] with the standards in the human capital accumulation literature. Past studies aiming at quantifying the ECRs wage returns (often called "salary premium") are mainly based on anecdotal evidence from interviewees' opinions and different methodological approaches (Bastianin and Florio, 2018; Florio et al., 2016; Battistoni et al., 2016). A few other studies, such as Camporesi et al. (2017) and Catalano et al. (2021), attempted to estimate Mincerian-like functions by including methodological artefacts to adapt econometric models to pre-existing survey data. However, these adaptations were not well-suited to accurately estimating wage returns from research training at RIs (see Section 2.3). We enhance the methodology by developing a research strategy specifically designed to address some of the challenges encountered in previous studies. Our strategy applies to any RI offering ECRs training opportunities. Additionally, our research lays the foundation for future studies aiming to employ more robust evaluation methods, including counterfactual design techniques.

[4] We use the terms "Big-science" and large-scale research infrastructures (RIs) interchangeably.

Second, we explore whether the wage returns from spending a research period at RIs are driven by an actual increase in productivity or by a reputational effect in line with the screening hypothesis. This approach is novel within the literature on the socioeconomic impact of RIs. Since Becker's (1962) seminal work, economists have widely accepted that increased investment in education and training enhances individuals' skills, thereby raising their productivity and, consequently, their earnings. In the early 1970s, Arrow (1973) and Spence (1973, 1974) introduced signalling models and argued that the positive relationship between education and earnings may not necessarily stem from improved productivity. Instead, they proposed that education serves as a signal to employers, providing information about individuals' innate abilities, a theory known as the screening hypothesis (Karasek and Bryant, 2012; Stiglitz, 1975). Similarly to education, training in world-class research environments such as RIs may reflect higher human capital rather than causing it in line with the screening hypothesis (Stiglitz, 1975; Arrow, 1973).

In addition to being methodologically intriguing, disentangling the productivity versus pure reputation effects has different research policy implications. If investing in RIs plays a major role in increasing ECRs' competencies and boosting productivity, the societal benefits from this pathway may contribute to justify public investment support alongside the scientific case. Conversely, if RIs primarily function as signalling tools, the case for public investment might be less compelling. Human capital accumulation for researchers is one of the key arguments for underpinning public funding of RIs (ESFRI, 2021; European Commission, 2020; OECD, 2019; Giffoni and Vignetti, 2019). In scenarios where RIs contribute to both human capital formation and signalling, it is important to assess the relative significance of each effect to evaluate their impact accurately. From this perspective, our work addresses an underresearched area in the economic evaluation of science, with implications for science policies. While the private returns to formal training programmes, e.g. in firms, are well-documented (Arellano-Bover, and Saltiel 2024), evidence of the human capital impact from public investments in (large-scale) RIs remains scarce.

The rest of the paper is structured as follows. Section 2 describes RIs as human capital incubators. We review the literature by discussing the state-of-the-art of examining wage returns to training at RIs vis-à-vis the reference economic literature linking training, work experience, and wages. Section 3 provides the theoretical and empirical framework and puts forward our research hypothesis. We also detail how this paper methodologically improves previous applications. Section 4 reports the results. Section 5 concludes with a discussion of the methodological and policy implications of our findings. It also reports the limitations of our research and suggests directions for future analysis.

# 2. Literature review

## 2.1 RIs as an avatar of human capital accumulation

The literature recognises the importance of RIs in delivering training to ECRs (see Catalano et al., 2021 for a literature review). This is often achieved through the cooperation of RIs with research-performing and higher-education institutions/universities via specific training courses, internships, scientific visits, and participation in defining and implementing academic curricula. There are many approaches (economic method, social approaches, etc.) that differ in the way the human capital impact of RIs is defined, measured and understood (European Commission, 2020; Giffoni and Vignetti, 2019). In this paper, we focus on microeconomic evaluation targeting individuals to study the relationship between wage returns, training at RIs, and work experience.

RIs play a crucial role in increasing the human capital of ECRs, making them more competitive and valuable in the labour market via multiple channels. These include access to cutting-edge resources and technology often unavailable in standard academic settings. Exposure to advanced tools enables ECRs to develop specialised technical skills and knowledge that are highly sought after in both academic and industrial sectors. Many RIs maintain links with industry, offering young researchers opportunities to work on applied research or industry-driven projects. This exposure deepens their understanding of market-oriented research and innovation processes, increasing their appeal to private-sector employers. Direct involvement in projects and experiments within RIs allows ECRs to apply theoretical knowledge to real-world problems, developing technical, scientific, and language skills, as well as independent thinking and critical analysis (Lu et al., 2023).

In addition to technical expertise, RIs provide ECRs with opportunities to develop crucial soft skills, such as project management, leadership, science communication, and problem-solving capacities, through collaborative, interdisciplinary learning and engagement with international networks.[5] Deming (2022) synthesises insights about human capital since Becker (1962) into four stylised facts, arguing that one of these is that skills like problem-solving and teamwork are increasingly economically valuable but that "the technology for producing them is not well understood" (Deming, 2022: 90). RIs often convene experts from diverse disciplines and regions, enabling ECRs to participate in interdisciplinary projects, enhancing their teamwork, communication, and networking skills – attributes highly valued across many sectors. Furthermore, many RIs maintain international partnerships, providing ECRs with opportunities to engage with global research communities, enhancing their visibility and prospects in the international job market, where cross-border collaborations are frequently essential (Karaca-Atik et al., 2023; Dusdal and Powell, 2021).

While the above set of skills is expected to increase ECRs' productivity, participation in high-profile RIs and large-scale projects also improves ECRs' credentials and recognition via, for instance, enhanced CV. The rationale for this reputational effect is that spending time conducting research at RIs is associated with higher earnings – not necessarily because it directly raises productivity but because it signals particular individual qualities to prospective employers. These may include scrupulousness in completing tasks successfully, strong self-motivation and drive, emotional maturity, and the ability to understand and internalise complex information and concepts (Bianchin et al., 2019).

## 2.2 The reference literature to study the link between earning, education, training and work experience

This section reviews the benchmark economic literature to assess the impact of training on earnings, to apply it to evaluate the human capital impact of RIs. Mincer (1974) starts from the economic theory of optimising behaviour and develops a model where identical individuals $(i)$ make forward-looking investments in human capital to maximise the present value of future earnings. Earnings functions à la Mincer are among the most applied models to study the determinants of individual wage, including education, working experience, and other individual characteristics.[6] Specifically, the model shows that the more individuals invest in education, the more they acquire skills that increase their productivity and, consequently, their earnings. The author derives a testable relationship known as "Mincer equation" (Eq. 1):

$$\ln wage_i = \beta_0 + \beta_1 EDU_i + \beta_2 EXP_i + \beta_3 EXP_i^2 + \varepsilon_i \quad (1)$$

[5] In Bloom's (1956) taxonomy of educational objectives, abilities such as teamwork and problem-solving are classified as "higher-order skills". The author outlines a hierarchical framework, with factual knowledge forming the foundation. This is followed by skills such as pattern recognition and classification, progressing to more advanced objectives like applying knowledge to novel situations, experimentation, connecting with new ideas, evaluation and decision-making, as well as the design and creation of new concepts.

For the individual $i$, the equation models (log) annual wages (or sometimes hourly wages) as an additive, linear function in the number of years of formal schooling, including tertiary education ($EDU$), and quadratic in the number of years of experience in the labour market ($EXP$).[7] The term $\varepsilon$ denotes the usual error term assumed to be independent and identically distributed (*i.i.d.*) among individuals and uncorrelated with both education and experience. The Mincer equation has become ubiquitous, to the point where people no longer reference the original source or may not even use the term "Mincerian" when studying earnings (Lemieux, 2006).[8]

In recent years, a significant body of research has examined the scope and impact of training, benefiting from newly available datasets that directly measure training activities (Adda and Dustmann, 2023; Grosemans et al., 2017). This research generally supports the human capital theory's prediction that a worker's wages tend to increase with prior investments in their training. Barron et al. (1989) highlight that "training is one of the few factors influencing both wage progression and productivity growth" alongside education. Similarly, Adda and Dustmann (2023) investigate the sources of wage growth over the life cycle, showing that it is determined by vocational training at the start of a career, unobserved ability, sectoral and firm mobility, and other factors, including cognitive-abstract or routine-manual skills. At the empirical level, basic specifications enrich the Mincer equation with training variables (Eq.2):

$$\ln wage_{ij} = \beta_0 + \beta_1 EDU_i + \beta_2 EXP_i + \beta_3 EXP_i^2 + \beta_4 T_i + \varepsilon_i \quad (2)$$

where $T$ denotes training. Measurements of training (hours of training, binary variable 1/0, and others) and the functional form of the relationship that links training to wage vary depending on the data and the hypotheses researchers want to test. As above $\varepsilon$ is a mean zero error term, uncorrelated with $T$. Recent research has suggested incorporating higher-order terms for experience and non-linearities in education and training, as well as other individual socioeconomic traits (e.g. gender, age, marital status, and many others). Although the introduction of these amendments, the Mincer equation has largely stood the test of time.

[6] The Mincerian function is used to examine earnings determinants; instead, other approaches can be applied to predict future salary trajectories over time.

[7] The logarithmic transformation of wage – whether annual or hourly – is typically applied to handle its skewed distribution and to meet normality assumptions. Regarding measuring education, sometimes different indicators for the different degrees of education is used instead of years of formal education. Working experience is typical defined as age minus school-starting age minus the number of standard years for completing each educational level. When estimating earning equations, EXP is typically modelled as a second-degree polynomial. This approach reflects the expectation that wage differentials linked to experience generally exhibit decreasing returns, as wage growth tends to be higher at the beginning of a career compared to later stages (Sørensen and Vejlin 2014).

The coefficient $\beta_1$ in Eq. (2) captures the wage return to an additional year of formal education (schooling). A literature review going back 60 years by Psacharopoulos and Patrinos (2018) based on 1,120 estimates in 139 countries from 1950 to 2014 suggests that an additional year of schooling increases wages by 10-12% percent (see also Deming, 2022; Gunderson et al., 2020).[9] The coefficients $\beta_2$ and $\beta_3$ on experience provide insights into how earnings evolve with increasing experience, capturing both the value of accumulated experience and the potential for diminishing returns as one's career progresses. Estimates of $\beta_2$ ranges from 0.04 to 0.10, while negative values of $\beta_3$ corroborates the diminishing returns hypothesis (Adda and Dustmann, 2023; Polachek, 2007). As regards training, the variable we are mainly interested in, studies suggest that formal training can lead to wage increases of around 3-6% annually on average, though this figure varies by industry, occupation, the quality of the training itself, and estimation procedures making comparisons difficult (Muehler et al., 2007; Frazis and Loewenstein 2005; Schone, 2004).

Applied economists have long acknowledged the complexity of estimating wage returns to training, largely due to the non-random nature of people receiving training. Selection into training programmes often correlates with unobserved individual abilities and personal traits (motivation, ambition, learning attitude, adaptability and openness to new experiences), which also impact earnings. As a result, simply regressing earnings on training does not accurately capture the causal effect of training on wages because $\varepsilon$ correlates with $T$, leading to overestimation of wage returns to training (Adda and Dustmann 2023, Frazis and Loewenstein, 2015; 2005; Brunello et al., 2012; Schone, 2004).[10] To make an analogy with RIs, ECRs' ambition may drive the decision to spend a period of training at a highly renowned RI, expecting better job opportunities, including higher wages in the labour market once the training is completed. To address the issue, empirical research has tried to control for time-invariant unobserved heterogeneity in earnings using fixed effects estimators when longitudinal panel data are available, under the assumption that innate ability does not change over time (Frazis and Loewenstein, 2006; 2005). Counterfactual techniques, which would allow for estimating a causal effect, such a regression discontinuity and diff-in-diffs estimators, have been less commonly used so far because of data limitation and challenges to finding control groups (see, e.g., Muehler et al., 2007). Similarly, instrumental variable methods are difficult to implement due to challenges in finding valid exclusion restrictions (Lee, 2009; Acemoglu and Pischke, 1999), namely finding an observable variable linked to the (decision of) training but uncorrelated with wage.

[9] Actually, literature reports different conclusions for different countries, but they are also sometimes mixed for the same country. Estimates of an additional year of education ranging between 6 and 18 percent, with a median in the 10–12 percent range. Across all OECD countries, the median earnings premium for a four-year college/tertiary education is 52 percent, or roughly 13 percent per year of education. See Based on OECD.Stat data at https://stats.oecd.org/Index.aspx?DataSetCode=EAG_EARNINGS.

[10] To mention jus a few, Schøne (2004), using Norwegian data, finds that after accounting for selection effects, the wage return to training is roughly 1 percent. Frazis and Loewenstein (2005) estimate the rate of return to formal training in the United States, noting that training raises wages by around 3 to 4 percent. Booth and Bryan (2005) analyse the impact of employer-funded training on wages in the UK, finding it increases wages by nearly 10 percent. Conversely, Leuven and Oosterbeek (2008) find no significant wage effects from training in Dutch firms once selective entry into training is accounted for. Muehler et al. (2007) examines the wage effects of continuous training programs using individual-level

## 2.3 The state-of-the-art of examining the effects of RIs on ECRs' human capital

This section discusses the existing microeconomic literature linking research training at RIs to ECRs' wages in relation to the above theoretical framework. To the best of our knowledge, this literature includes only a handful of studies, which remain valuable points of reference despite some methodological caveats.

Florio et al. (2016), further elaborated by Florio (2019), aimed to estimate ECRs' wage returns (referred to as the 'salary premium') from spending a period conducting research at the CERN Large Hadron Collider (hereafter, LHC). The study was based on a survey targeting 385 students at doctoral level, comprising both ECRs still at CERN during the survey period (40% of the sample who had not yet entered the labour market) and former ECRs who had completed their "hands-on" working and training experience in one of the experiment detector projects associated to the LHC and were active in the labour market (60%). The salary premium was derived from the following question: "To what extent do you expect that your future salary will be higher than that earned by somebody else?" with options: 0%, up to 5%; 5-10%; 11-20%; 21-30%; more than 30%. The weighted average of the responses yielded a salary premium of 9.3%, with similar averages reported for both current and former ECRs. Additionally, the authors assumed "an additional small premium of 2-3% due to the composition effect of job opportunities across occupations" (Florio et al. 2016:12), resulting in a combined estimate of an 11.8% wage annual premium. Florio et al. (2016)'s results were used by data from the German Socio Economic Panel (GSOEP). Using the counterfactual DID estimator, they find a wage effect of about 5–6. Almeida-Santos et al (2010), using British data, observe that training is associated with wage dispersion only among white-collar workers. Görlitz (2011), drawing on German employer-employee data, examines the effects of on-the-job training considering also different number of courses attending by employees. The author finds that attending a single training course does not significantly impact wages. Lastly Picchio and van Ours (2013) use Dutch data and find that firm-provided training significantly increases future employment prospects.

Bastianin and Florio (2018) to attribute a salary premium to ERCs at CERN High-Luminosity LHC (HL-LHC), a major upgrade of the LHC machine. Battistoni et al. (2016:87) also relied on Florio et al. (2016) to calculate the salary premium accruing to ECRs at the CNAO, the Italian National Hydrotherapy Centre for Cancer Treatment.

The single-question approach provides a direct insight into the perceived wage premium, reflecting respondents' own expectations. However, the methodology presents a number of caveats. The authors aimed to approximate a counterfactual analysis, but the design did not fully adhere to established counterfactual methodologies (Angrist and Pischke, 2009). Respondents were tasked with processing multiple layers of complex information. For instance, to respond to the survey question, current ECRs were required to (i) predict their long-term future salaries; (ii) estimate the long-term salaries and career trajectories of their peers (defined as individuals not accepted into CERN-related programmes); and (iii) calculate the percentage difference. These tasks required significant cognitive effort and assumed a high degree of foresight and comparative knowledge among participants. The authors argued that former ECRs, having gained firsthand knowledge of job market opportunities, could compare their expectations with those of their peers. While this rationale is reasonable, it also assumes that respondents possess detailed insights into the prospective careers of peers who were not part of the CERN experience. These peers could come from universities worldwide and enter diverse sectors globally after their studies, making accurate comparisons challenging. Consequently, while the question captures valuable opinions, the feasibility of such estimations raises concerns about potential bias in the derived salary premium.

The authors corroborated the 11.8% salary premium figure by situating it within the range of returns to higher education reported in the literature, citing, e.g., Montenegro and Patrinos (2014). Florio (2019: 177–118, 122–124) further contextualised the estimated premium by comparing it with the average return on tertiary education, including PhD, estimated in various countries and periods. While such comparisons are legitimate, their appropriateness calls for closer examination. Schone (2004) highlights that when estimates of returns to training align closely with those of education, this warrants further scrutiny because the literature on wage returns to education (including university education) and wage returns to training constitutes two distinct streams of research with differing findings. Given the typically shorter duration of training programmes, he attributes such anomalies to unobserved factors influencing wage levels. Moreover, the CERN related salary premium discussed in Florio et al. (2016) and further analysed by Camporesi et al. (2017) and Catalano et al.

(2021) (see below) refers to an average training duration of 44 months (3.6 years).[11] In comparison, the average earnings premium for a four-year college/tertiary education across all OECD countries lies between 48-52% (Deming, 2022).[12] In the context of the US job market, Florio (2019: 123) compares the starting salaries of individuals with a PhD to those with a master's degree , estimating a "PhD salary premium" of 34%. This figure significantly exceeds the wage returns associated with practical training in a scientific research project at CERN, and this is because wage returns to training are typically lower than those to formal education, as discussed in Section 2.2. On top of that, making direct comparisons is challenging due to measurement issues inherent in the various typologies of training and delivery methods. These challenges are further compounded by the interplay between theoretical frameworks and empirical approaches in wage-training research (Black et al., 2023). The underlying mechanisms driving wage returns – such as the nature of acquired skills or the delivery mode of training – differ significantly between formal education and training (Adda and Dustmann, 2023; Black et al., 2023; Grosemans et al., 2017; Frazis and Loewenstein, 2015). OECD (2014) posits that the intellectual environment within scientific research projects, such as those implemented at CERN, is akin to that of the most advanced high-technology firms. This environment may differ markedly from formal tertiary education, making the literature on wage returns to training a more relevant benchmark for comparison.

Finally, it is important to consider that ECRs' expectations of future salaries might be influenced by overconfidence following their training experience. Research suggests that overconfidence can lead to inflated salary expectations after graduation (Schnusenberg, 2020), potentially resulting in overestimating the salary premium. Furthermore, such expectations may introduce measurement errors in eliciting perceived wage returns (Serrano and Nilsson, 2022).

Camporesi et al. (2017) exploited all the survey data that Florio et al. (2016) collected and attempted, for the first time, to estimate a Mincerian-like equation in the field of RIs. In addition to the salary premium question discussed above, the survey also collected information on the annual (expected) salary category of the (current) former ECRs at the LHC. Salary categories were: < EUR 30 000; 30 000 – 40 000; 40 000 – 50 000; 50 000 – 60 000, and > 60 000. This categorical variable was regressed on a set of respondents' personal characteristics (gender, age, being employed, nationality), career-related information (sector of employment) and information about the presence at CERN (duration, type of skills acquired). An ordered logistic regression was run to fit the categorical nature of the salary question.

[11] See Camporesi et al. (2017: Table 2, p.11) and Catalano et al. (2011:14).

[12] This corresponds to about 12-13% per year of tertiary education (as discussed in Section 2.2). Evidence in Deming (2022) is based on OECD.Stat data at https://stats.oecd.org/Index.aspx?DataSetCode=EAG_EARNINGS.

The study identified the duration of the presence at CERN as a key factor influencing the probability of respondents declaring higher salary categories. Specifically, Camporesi et al. (2017:19) documented that for a respondent with an average research duration of 44 months, the probability of reporting a salary in the EUR 50,000–60,000 range was 5% higher than declaring a salary below EUR 30,000, which served as the reference category. This probability increased to 12% when considering salary declarations exceeding EUR 60,000. These findings provide useful insights into the determinants influencing declared salary categories; however, the study does not estimate a "salary premium" in the sense of counterfactual comparisons as outlined in the paper.[13] Furthermore, since Camporesi et al. (2017) relied on the same dataset as Florio et al. (2016), the analysis inherits several limitations previously discussed regarding the reliability of information reported by current ECRs. Additional concerns arise from the omission of key explanatory variables – such as prior work experience and career-related factors – needed to accurately estimate Mincer-type equations like Eq. (2). The absence of these variables raises the possibility of omitted variables bias, potentially inflating results by combining the effects of CERN experience, formal education, and career development. Another notable issue relates to sample composition. Camporesi et al. (2017) included all respondents aged up to 44 years (n = 318) in their analysis without isolating ECRs from more experienced professionals (see Table 2:11). This decision further complicates the interpretation of results, as differences in respondents' career stages may confound the observed effects.

Catalano et al. (2021) sought to expand the analysis of the "salary premium" by re-launching the survey used in Camporesi et al. (2017). They collected 438 valid responses, extending the sample size, and repeated the multivariate analysis. As an additional contribution, the authors administered a second survey targeting team leaders of ECRs – senior scientists affiliated with universities or research institutes instrumental in enabling ECRs to pursue research opportunities at CERN. The second survey introduced a novel perspective by eliciting opinions on the reliability of salary expectations reported by ECRs in the earlier study. Each institute participating in an experiment or project at CERN is required to appoint a Team Leader and a Deputy Team Leader responsible for making sure that all the members of their team are aware of CERN's regulations and requirements and of their duty to comply with them throughout their stay at CERN. This person is an employee of the participating institute. She/he administers persons that are sent by the institute to perform research at CERN for limited periods of time. One question asked team leaders to assess whether a salary premium ranging from 4% to 12% was reliable. Based on 322 valid responses, the distribution of answers was as follows: “the range sounds reasonable to me” (54%), “I would have expected a greater impact” (31%), “I would have expected a lower impact” (2%), “I have no opinion” (1%), and “I do not know at all” (12%). The way how the question was formulated raises several methodological concerns. Research in behavioural economics highlights that presenting predefined ranges can influence respondents’ answers by anchoring their expectations to the suggested values, thus introducing cognitive biases (Furnham and Boo 2011; Chapman and Johnson, 1994). Such biases may lead participants to unconsciously conform to perceived expectations, undermining the validity of the results. Techniques to mitigate anchoring effects – such as providing open-ended questions or varying the order of options – were not implemented in Catalano et al. (2021). Consequently, the analysis relies heavily on team leaders’ perceptions without sufficient safeguards against potential distortions. An alternative approach to validate the salary expectations expressed by ECRs might have involved presenting team leaders with the same set of salary ranges as those used for ECRs (0%, up to 5%; 5–10%; 11–20%; 21–30%; more than 30%). Such alignment would have enabled direct comparisons between the two groups’ responses and tested the assumption that team leaders possess sufficient information about ECRs’ career trajectories to evaluate salary expectations reliably. [14]

[13] Camporesi et al (2017: abstract) state that the study documents a "*LHC salary premium' ranging from 5% to 12% compared with what they would have expected for their career without such an experience at CERN*."

Building on these insights, this paper aimed to design a research strategy for estimating wage returns associated with research periods at RIs using tools grounded in human capital theory discussed in Section 2.2.

# 3.Method

## 3.1 Working hypotheses

As an international laboratory, one of CERN’s goals has been to bring scientists together across national boundaries and train students in a global environment. Considering the period 2010 – 2019, on average, each year, CERN hosted more than 10,000 Users and Associated Members of the Personnel (MPA) from hundreds of institutions across the globe, as well as 180 doctoral students in the frame of CERN’s own doctoral student programme (FCCIS, 2024).[15] Users and MPAs are scientists, researchers, engineers and technicians from universities, research institutions, and laboratories worldwide participating in various research projects. They are not CERN staff members but contribute to projects at CERN. Examples include particle accelerator projects, experiment detector projects, computing and software development activities. Users and MPAs are physically at CERN for limited periods of time. They are typically not continuously at CERN. Stays range from weeks to months. About 50%, however, stay for longer periods of time up to three years. They also perform work related to the project at CERN at their home institutes. During this period, in particular ECRs, perform a wide range of activities related to the research projects, including designing and building detectors, developing software and hardware, running experiments, analysing data, and producing scientific and technical documents. They also work embedded in large, international teams, contributing to the operation of large-scale experiments (e.g., ATLAS, CMS, LHCb, ALICE), particle accelerators (LHC, PS, SPS, LEIR, ELENA, AD) or supporting developments in accelerator technology and computing.[16]

[14] Alternatively, the authors could randomly submit different ranges of salary premium to different sub-samples of Teams leaders and see if 4-12% received more accordance than others as practises in contingent valuation studies (Giffoni and Florio, 2023).

[15] In the period 2010 – 2019, on average, each year, CERN counted about 2,500 staff (with both an indefinite and limited contract), 850 Fellows and Associates, more than 10,000 Users and 180 doctoral students. In this paper we are mainly interested in the last two categories, with the Users being the bulk of the ERCs.

In parallel, CERN has, among a set of training programmes, a doctoral student programme that offers students after their Master's degree the opportunity to carry out practical work of their studies at CERN, embedded in a project for a maximum period of up to three years. Doctoral students remain affiliated with their home universities from which they eventually obtain their degree but carry out part or all their practical work at CERN under the supervision of a CERN employee. The programme complements their university studies, allowing students to benefit from CERN's cutting-edge facilities and the fertile, cross-sectoral environment. Research topics typically align with CERN's focus areas, such as applied physics, many different technical STEM disciplines, information technology, law, international business administration, economics, environmental engineering and others. They are embedded in CERN's vibrant international community, where they can attend seminars and conferences and collaborate with scientists and engineers worldwide. Doctoral students receive a subsistence for stays up to 24 months that can under exceptional conditions be extended to 36 months.

Our first two hypotheses suggest that spending a period of research training at CERN increases the skills of ECRs and, in turn, generates better job opportunities (Bianchin et al, 2019),

[16] For instance, superconducting cables and magnets, radiofrequency systems, radiation tolerant and hard electronics, data processing software, data transmission technologies, collaborative software platforms, safety systems, environmental studies, knowledge transfer and an ever-growing number of activities related to the administration, organisation and performance measurements of large science programmes and projects.

including wage-related effects via increasing productivity as people acquire competencies which would be otherwise difficult to obtain elsewhere in the labour market (OECD, 2014):

| | |
|---|---|
| H1. | Training in a project at CERN impacts on ECRs' wage |
| H2. | The skills acquired by ECRs determine the impact of the active period at CERN on the wage |

Participating in high-profile RIs and large-scale projects improves ECRs' credentials and recognition via, e.g. enhanced CVs. Accordingly, training at RIs may reflect higher human capital rather than causing it to be in line with the screening hypothesis. Put it differently, CERN is associated with higher earnings after the student moves to another employment, not because it directly raises productivity, but because it signals certain individual characteristics to prospective employers. These characteristics include scrupulousness in completing tasks, strong self-motivation and drive, emotional maturity, and the capacity to grasp and internalise complex information and concepts. This leads to our third working hypothesis:

| | |
|---|---|
| H3. | Training in a project at CERN influences ECR's job opportunities through reputation mechanisms, complementing its effects on productivity |

## 3.2 The survey

To test the hypotheses, we designed a dedicated survey and structured the questionnaire in three main sections, bearing, respectively on: (A) "personal information and education background"; (B) 'your current job", and (C) "experience at CERN and its impact on your career". Table 1 reports a synthetic version of the questionnaire where the main variables are visualised (see Annex A for the full questionnaire).

Table 1: Synthetic version of the questionnaire

| Section A: Personal Information and background | Section B: Current job | Section C: Experience at CERN and impact career |
|---|---|---|
| • Nationality and residence<br>• Gender<br>• Date of birth<br>• Marital status<br>• Type and level of formal education<br>• Type and number of courses beyond formal education<br>• Only for PHDs: name of university and dept | • Current occupational status<br>• N of years in the labour market<br>• Type, sector, and size of firm<br>• N of weekly working hours<br>• Yearly gross salary<br>• Type of contract<br>• Extra benefits | • Start and (expected) end date at CERN<br>• Domain of activity<br>• Considerations on the importance to apply for a period a CERN<br>• Share of working time at CERN spent on different activities<br>• Skills improvement |

Source: authors

The survey targeted people who were formerly carrying out practical work at CERN in a research project and who were employed elsewhere at the time of the data collection. The survey was managed by CERN staff, who distributed it via direct emails to researchers in their database and administered it using the Drupal software. The on-line survey was run between April 2020 and September 2022. In total, 2,600 individuals were reached, and 710 answers were collected, with a response rate of 27%. For privacy reasons,[17] we received the anonymised database of answers and performed a data-cleaning procedure to identify eligible candidates for our analysis. We identified a sample of 196 valid individual responses (Table 2).[18] Unlike earlier studies, we allow ECRs into our sample only after they have completed their training to collect information on their actual wage. Table 3 shows the main differences between our survey and the surveys employed in the above-mentioned previous studies.

Consistently with the literature on skills formation in the transition from university to work, we focused on young researchers under 40 years old. The literature highlights the importance of age in the returns to training for future career paths and labour market outcomes, including wage levels during the early stages of a career (Grosemans et al., 2017; Evers and Rush, 1996). On top of that, we only retrieved ECRs who had spent no longer than eight years at CERN. Activities in a high-tech research project turn out to be a human and social capital incubator if people have time to acquire the knowledge, skills and competencies that are highly valued in the labour market (Catalano et al., 2021; Bianchin et al., 2019). On the other hand, the cut-off of eight years at CERN reflects the maximum duration of a limited duration contract before receiving an open-ended contract, as laid out in the CERN HR Policy.[19] Accordingly, the threshold excludes people who remain at CERN and are out of the scope of the present study.

[17] Regulation (EU) 2016/679 of the European Parliament and of the Council of 27 April 2016 on the protection of natural persons with regard to the processing of personal data and on the free movement of such data, and repealing Directive 95/46/EC (General Data Protection Regulation) (Text with EEA relevance)

[18] The survey was supposed only to reach out to former CERN researchers. However, to increase the chance of obtaining a larger number of responses, in October 2022, 14 notices were posted on the Facebook Group Young@CERN (which has 16,100 users), and 7 posts were made on Alumni CERN (8,363 users). This dissemination process also reaches out to a number of CERN employees (staff) and people still doing research CERN. We excluded both categories from our analysis as they do not fit our research purpose.

Our cleaning procedure reduced the potential number of observations available for econometric analysis; however, it ensured that the target sample is appropriately defined, focusing on individuals for whom salary returns are most likely to materialise. Simultaneously, it preserved enough valid responses to enable econometric modelling and robust statistical inference. This represents an improvement over previous studies, which did not distinguish between subgroups within the sample. Robustness checks corroborate our main findings (Section 4.3).

Table 2 - Steps to identify the target sample

| # | Action | N. of respondents |
|---|---|---|
| 1 | Total number of respondents | 710 |
| 2 | Removal of duplicates or inconsistent records | 670 |
| 3 | Removal of responses with missing values of the main variables (i.e.., salary, time spent at CERN) | 549 |
| 4 | Exclusion of respondents who were still contributing to research projects at CERN at the time of the survey and CERN employees (staff) | 347 |
| 6 | Exclusion of respondents with a research training period above 8 years | 326 |
| 7 | Exclusion of respondents above 40 years old | 196 |

Source: authors' elaboration on survey data

Table 3 – Difference between the new survey and earlier surveys

| | New survey | Earlier surveys |
|---|---|---|
| Target | Former researchers who had already left the research projects at CERN and were in the labour market at the time of the survey | Both researchers who were still contributing to research projects at CERN at the time of the survey (40%) and former researchers in the labour market (60%) |
| Focus on ECRs | Stricter focus on sample selection: the target strategy, as reported in Table 2, has been more rigorous, considering only researchers under 40 years old, with a specific focus on the sub-sample of those under 35. | Weak focus. All the samples considered up to 44 years old. |

[19] Source: CERN HR Department.

| | | |
|---|---|---|
| Salary data | Actual salary earned in monetary terms | Expected (40%) and actual salary (60%) in predefined salary categories. Most of them were expectations regarding end-career salary. |
| Salary-related information | Hours of work, type of contract, recipient of extra-bonuses beyond salary, income from other activities (yes/no) | Not available |
| Experience in the labour market | Number of years of working experience | Not available |
| Education | Number of years of formal education | Level of education (PhD yes/no in the analysis) |
| Information on other training experiences in addition to CERN | Yes | Not available |

Source: authors

## 3.3 Empirical set up

In a human capital framework, ECRs face a decision to apply for a research period at CERN or pursue alternative paths, such as continuing their careers at home institutions, exploring opportunities in the broader labour market, or engaging in further education or training to enhance their human capital. Testing our working hypotheses requires establishing a statistically significant relationship between salary and research training time spent at CERN and, subsequently, between skills acquired and training time. Identifying a mechanism linking skill development, training, and salary outcomes would support the hypothesis that training contributes to salary increases via the productivity channel rather than solely through reputational effects (see Section 4.2.2).

We estimate the following earnings equation (Eq. 3):

$$\ln wage_i = \beta_0 + \beta_1\, EDU_i + \beta_2\, f(EXP_i) + \beta_3\, T_i + \beta_4\, f(CERN_i) + \varepsilon_i \quad (3)$$

where $i$ denotes the ECR, $CERN$ represents the duration of the training in months, and $f(.)$ is the functional form linking training duration to wages to be empirically determined (Frazis and Loewenstein, 2005). If $f = \ln(CERN)$, the parameter $\beta_4$ captures the linear effect of an additional month of training on wages. Alternatively, if $f(CERN) = \beta_4\, CERN + \rho CERN^2$ the quadratic term reflects diminishing marginal returns, where the parameter $\rho$ is expected to be negative, and the return is given by Eq. (4):

$$\frac{\delta wage}{\delta CERN}\beta_4\; 2 * \rho CERN \quad (4)$$

Independently of the functional form, the CERN wage return in Eq. (3) sums up to return to education ($EDU$), experience ($EXP$),[20] and other training beyond CERN ($T$).

An unbiased OLS estimation requires $\varepsilon_i$ to have a zero mean and to be uncorrelated with CERN, a condition met only if participation in CERN training is random. However, training participation is influenced by factors such as past academic performance, ambition, and unobserved individual abilities, which also affect wages. We control for observable factors that may jointly influence training duration and salary formation. Firstly, we account for observable confounding factors by estimating an augmented model (Eq. 5):

$$\ln wage_i = \beta_0 + \beta_1 EDU_i + \beta_2 f(EXP_i) + \beta_3 f(CERN) + \beta_4 \bar{S_i} + \beta_5 \bar{Z_i} + \beta_6 \bar{C} + \varepsilon_i \quad (5)$$

where $\bar{S_i}$, $\bar{Z_i}$, and $\bar{C}$ are vectors of salary-related information, individual traits, and contextual variables, such as country of employment and sector.[21] Subsequently, to test H1b and H2, the earnings function is further augmented to include skills acquired or improved during the CERN training as in Eq. (6):

$$\begin{aligned}\ln wage_i = {} & \beta_0 + \beta_1 EDU_i + \beta_2 f(EXP_i) \\ & + \beta_3 f(CERN) + \beta_4 f(Skills)_{ij} + \beta_5 \bar{S_i} + \beta_6 \bar{Z_i} + \beta_7 \bar{C} + \varepsilon_i \end{aligned} \quad (6)$$

where $f(Skills)_{ij}$ represents different functional forms, including the number of skills acquired or a binary variable indicating whether at least one skill was acquired. The subscript $j$ denotes the skill type (hard or soft). Following Camporesi et al. (2017) and Catalano et al. (2021), we grouped scientific and technical skills, critical analysis, problem-solving capacity, and independent thinking under the label "hard skills", while communication, team/project leadership, developing maintaining and using networks of collaborations under the label "soft skills" (see Section 4.2.1 for details). Section 4.2.3 also proposes a set of robustness estimates that partially account for ability bias in Eq. (5) and jointly control for it in Eq. (6).

[20] Despite usually considered in its quadratic form, when the working experience horizon refers to the first years of experience, it can be considered linear experience (Sørensen and Vejlin 2014).
[21] $\bar{C}$ is not individual-dependent and therefore the subscript $i$ is omitted.

# 4. Analysis and results

## 4.1 Descriptive analysis of the targeted sample

ECRs in our selected sample were from 91 different research institutions worldwide, most from Europe. Respondents were predominantly male, with 86% (n = 168) reflecting the male-dominated nature of the HEP research field (Holman et al. 2018). The age of respondents ranged from a minimum of 24 years (1%; n = 2) to a maximum of 40 years (3%; n = 6). Specifically, 59% (n = 116) were under 35 years old, and an additional 15% were 38 years old (Table B.1, Annex B).

The majority of ECRs (61%; n = 119) held or were pursuing a doctoral degree (Table 4), which is consistent with the high average number of years of formal education in our sample (19.8 years). More than half of the sample (51%; n = 111) had attended additional training courses alongside their formal education and CERN research training (Table B.1, Annex B).

Table 4 – Education level of respondents (*N = 196*)

| Education level | N | (%) |
|---|---|---|
| Doctoral degree | 112 | 57% |
| Doctoral degree ongoing (at the time of the survey) | 6 | 3% |
| Higher professional education | 6 | 3% |
| Master degree | 62 | 32% |
| University degree (below master's) | 9 | 5% |
| Secondary diploma | 1 | 1% |

Note: Rounded figures. Source: authors' elaboration

Through the survey, we investigated the research training experience at CERN from multiple perspectives, posing a series of questions that addressed various dimensions of the training process. These included the decision to apply, the duration of the training, the nature of the research activities undertaken, and the skills acquired or enhanced throughout the research experience. To gain a deeper understanding of the factors driving their decision to apply, we presented ECRs with a set of options and asked them to rank them according to their importance. The opportunity to work in an international environment and develop new skills emerged as the primary motivations for applying to CERN, with over 90% of respondents categorising these factors as "Important" or "Very Important." The desire to deepen specific knowledge ranked next, with 88% of respondents acknowledging its significance. In contrast, improving job prospects was cited by 60% of ECRs, while 12% considered it either "Not at all important" or "Low importance."

Figure 1. Motivation to apply to a research training experience at CERN (*N = 196*)

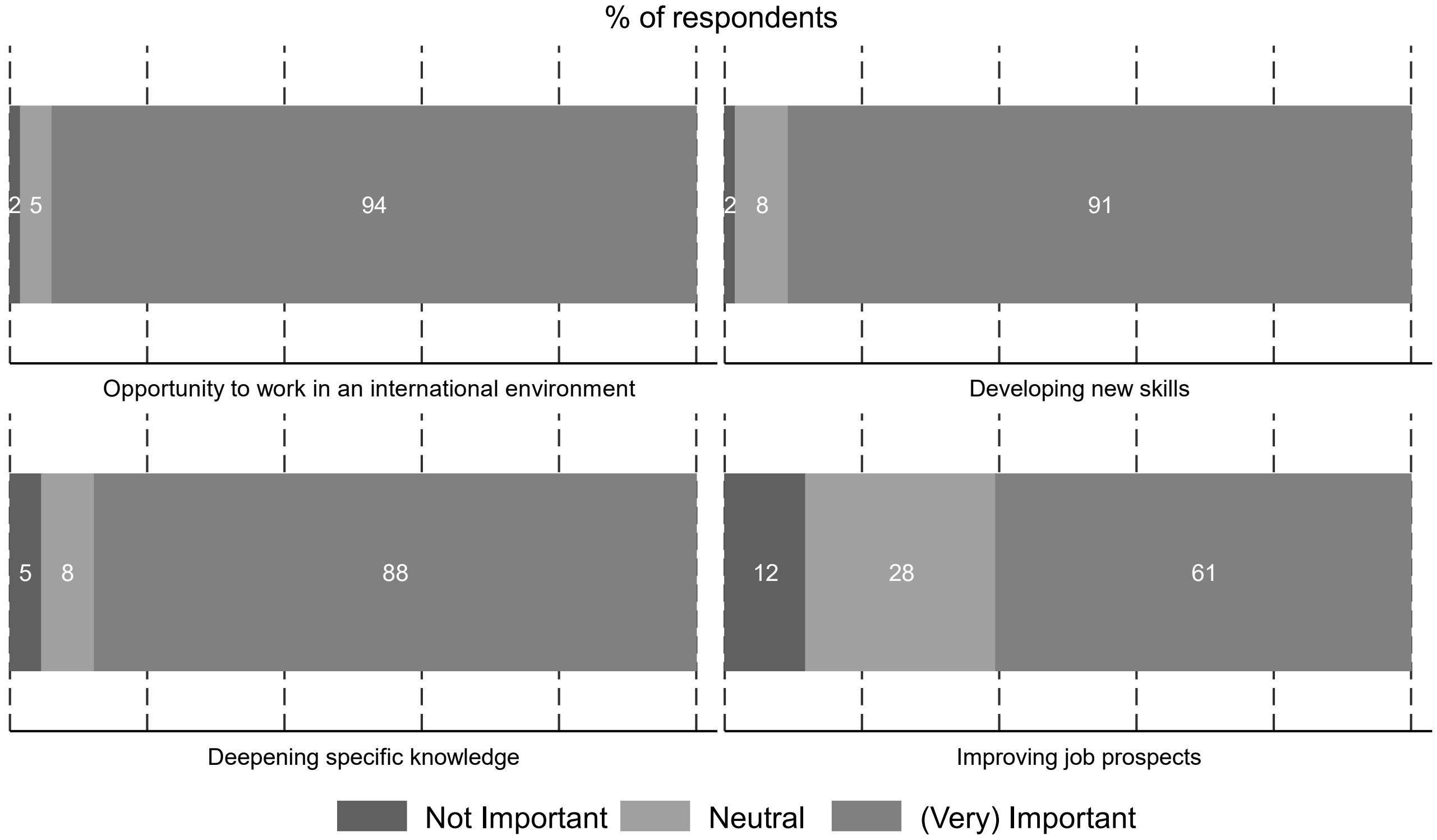

Source: authors' elaboration. Note*:* 196 valid answers analysed. Multiple answers question. Question C.4 of the questionnaire: "How do you rate the importance of the following considerations on your decision to apply for a working experience at CERN? Respondents were asked to indicate their answer on a Likert scale including 5 options (Not at all important, Low importance, Neutral, Important, Very important). In the chart, the option "(Very) important" groups the original options "Very Important" and "Important" while "Not Important" groups the original options "Low importance" and "Not at all important".

The duration of active period at CERN lasted between 1 and 5 years for 86% (*n = 169*) of them, with a peak of 2 years. The average duration in the sample is 3.4 years (Figure 2).

ECRs were mainly engaged in experimental physics (*32%; n = 63*), engineering (*28%; n = 54*), particle accelerator physics (*14%; n = 28*), and informatics (*13%; n = 26*) research domains. The remaining part carried out research or other activities in different domains, mostly including theoretical physics, management and administration (e.g. marketing, health and safety, human resources) (Figure 3).

Figure 2. Duration of the training experience at CERN before leaving (*N =196*)

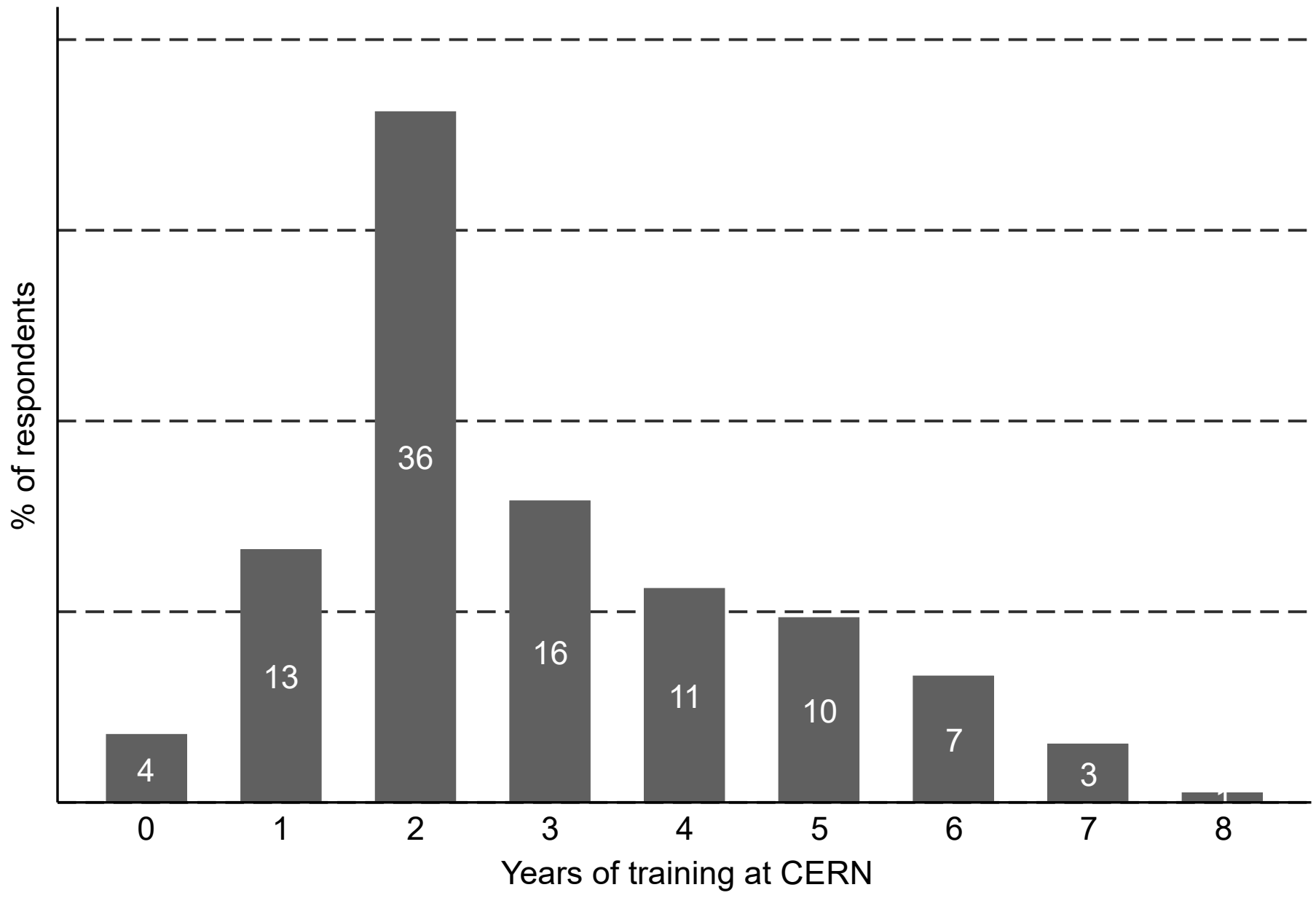

Source: authors' elaboration. Note*:* 196 answers analysed.

Figure 3. Domain of activity while working at CERN (*N = 196*)

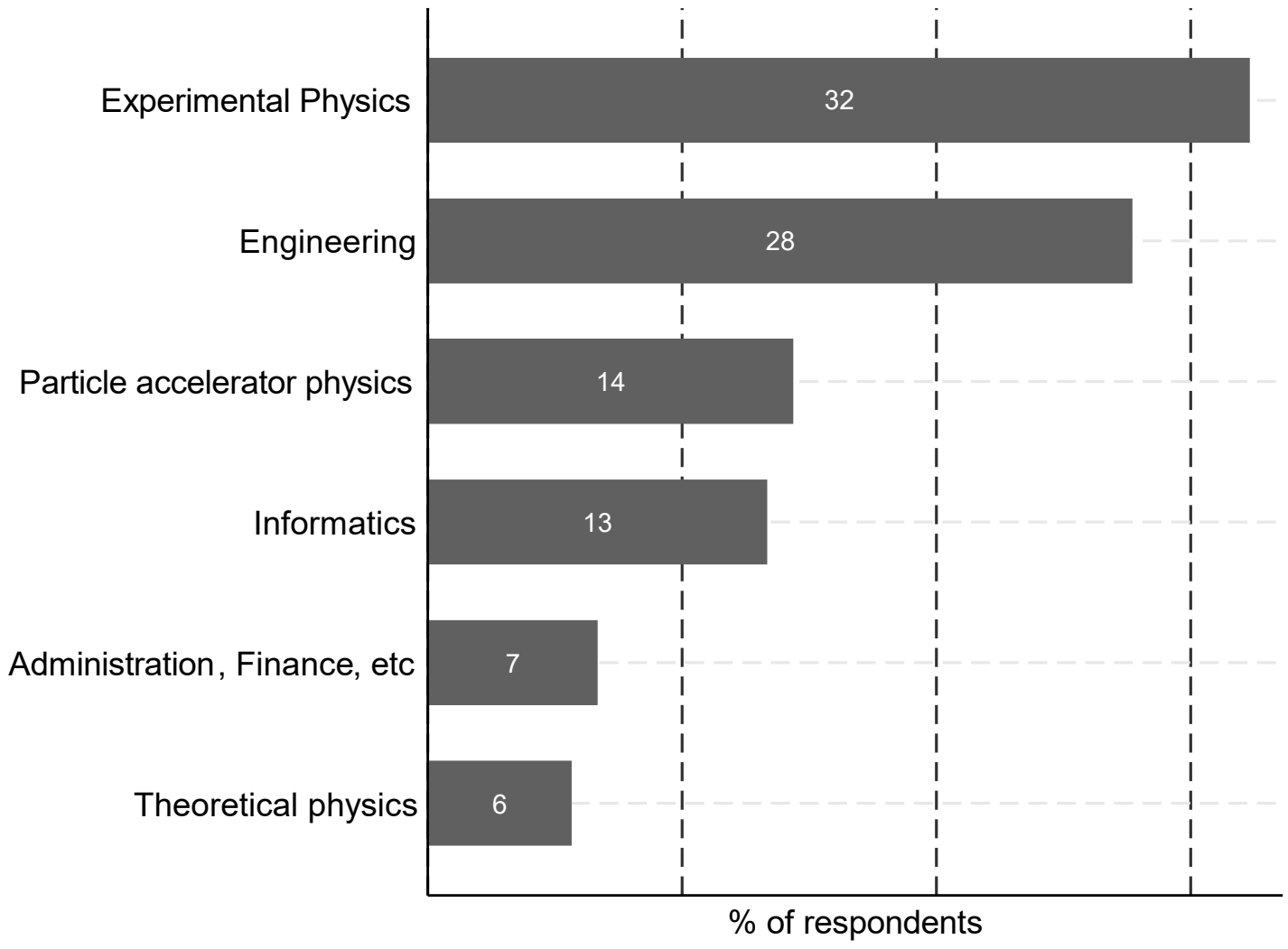

Source: authors' elaboration. Note*:* 196 answers analysed. Multiple answers question. Question C.3 of the questionnaire.

One of the key aspects of the survey has been collecting detailed information on skills. Measuring skills present challenges, and their predictive power on earnings may vary depending on the metric adopted. Our questionnaire followed the most typical approach, which involves using Likert scale items without any cardinal meaning. Technical competencies (e.g., software development and data analysis) and scientific and problem-solving capacity are the top three declared by ECRs in our sample. However, other skills such as language competencies, independent thinking, cultural, social, communication and leadership skills also emerge as important competencies developed by ECRs (Figure 4).

Figure 4. Competencies acquired or improved during the research training at CERN (*N = 196*)

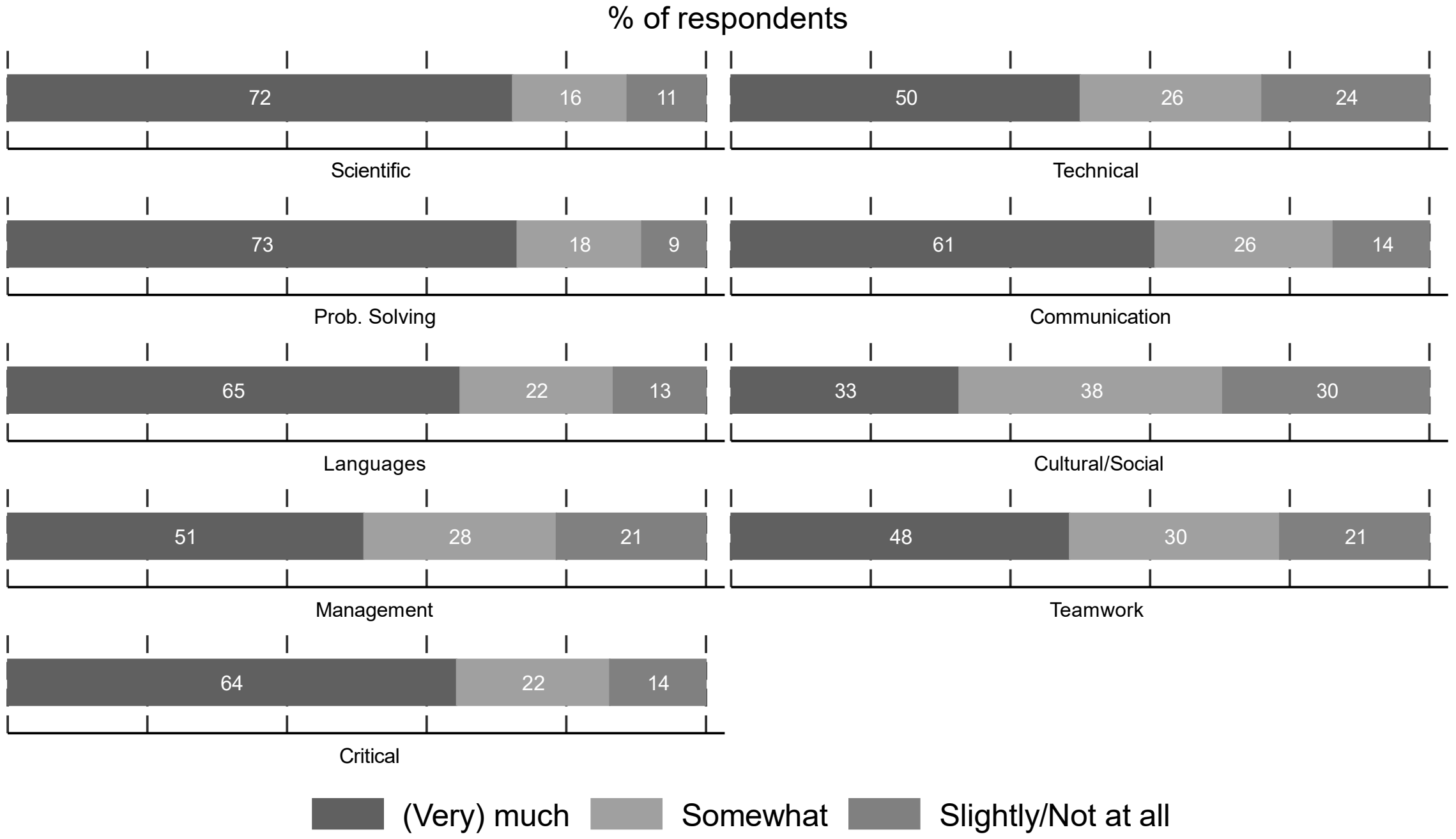

Source: authors' elaboration. Note*:* 196 answers analysed. Multiple answers question. Question C.5 of the questionnaire: "Indicate your skills improvement experience for each skill below. Thanks to my experience at CERN, I have improved my:". Respondents were asked to indicate their answer on a Linkert scale, including 5 options (Not at all, Slightly, Somewhat, Much, Very Much). In the chart, the option "(Very) Much" groups the original options "Very Much" and "Much" while "Not at all – Slightly" groups the original options "Slightly" and "Not at all". Refer to Table B.2 in Annex B for further details on the question items.

All the ECRs in the sample had occupations at the time of the survey, which were obtained after their period at CERN. The average number of years of working experience in the labour market was 4 years (Table B.1, Annex B), with 71% (*n = 140*) of respondents having a working career of less than 5 years. About 3% were in the labour market for more than 10 years and 2% for less than 1 year since leaving CERN (Figure B.1, Annex B). In terms of the employment sector, 50% (*n = 101*) of respondents were employed in the industry and finance, followed by academia and research (41%; *n = 101*), a finding in line with previous studies (Catalano et al., 2021; Bianchini et al., 2019) (Table 5, Panel A). Specifically, ECRs in engineering mainly found an occupation in the industry and finance sector, while physicists were almost equally distributed between industry and finance, and academia. Training in informatics also helped ECRs entered positions in public administration in addition to industry (Figure 5).

Figure 5. Relationship between domain of research activity at CERN and sector of employment

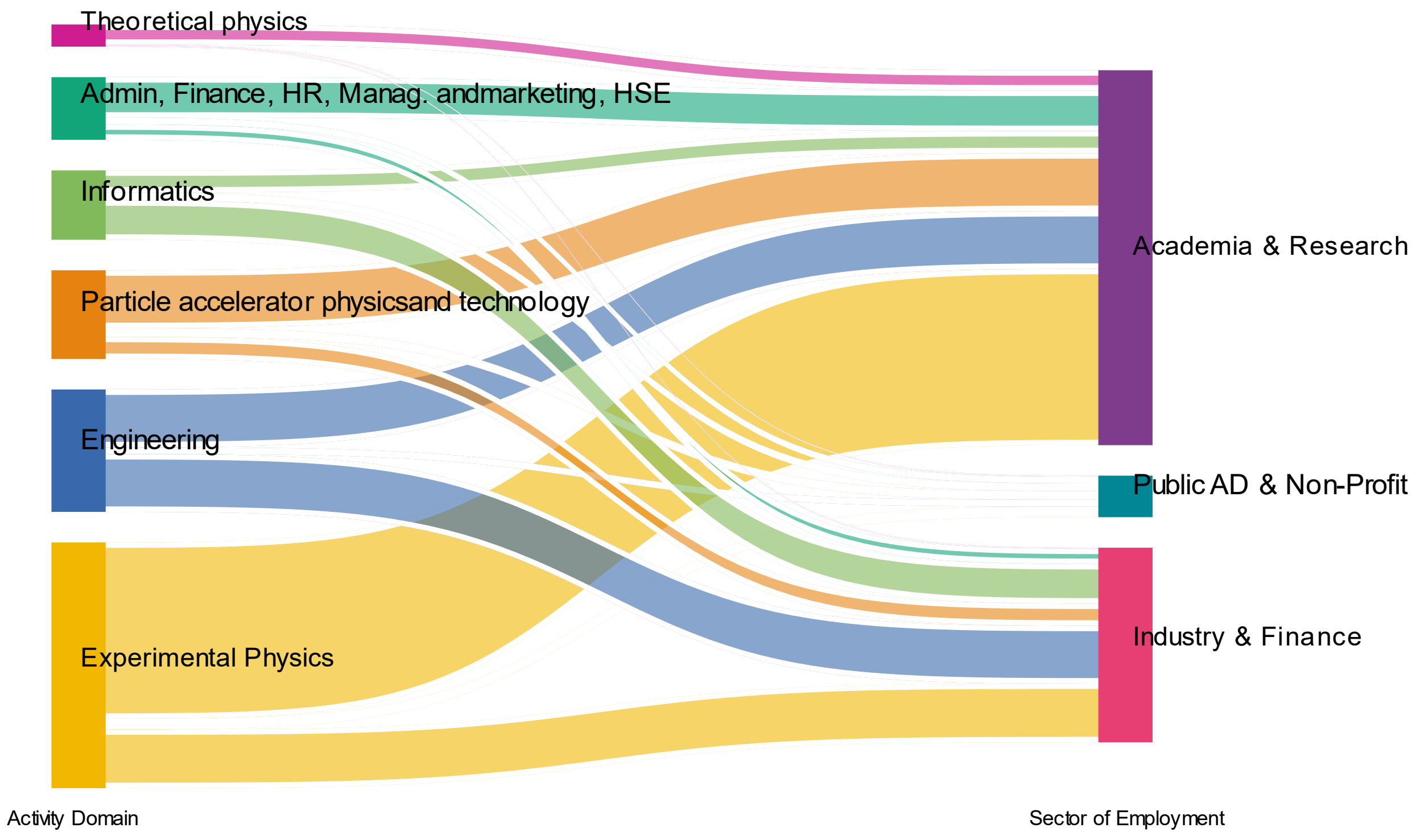

Source: authors' elaboration.

Large organisations and enterprises located in Switzerland, Germany, France, the United Kingdom, and the United States hired most of the ECRs, an evidence in line with other similar studies (e.g. Bianchin et al., 2019) (Table 5).

Table 5 – Occupation of respondents by sector, firm size, and country (*n = 196*)

| Panel A: sector of employment | N | (%) |
|---|---|---|
| Industry and finance | 99 | 50% |
| Academia and research | 80 | 41% |
| Public Administration, including non-profit | 17 | 9% |
| No answer | - | - |
| Panel B: occupation by organisation / firm size | | |
| Large | 148 | 75% |
| Medium | 25 | 13% |
| Small | 11 | 6% |
| Micro | 9 | 5% |
| No answer | 3 | 1% |
| Panel C: country of employment | | |
| Switzerland | 45 | 23% |
| Germany | 22 | 11% |
| France | 16 | 8% |
| United States | 15 | 8% |
| United Kingdom | 14 | 7% |
| Italy | 11 | 6% |
| Netherlands | 9 | 5% |
| Austria | 8 | 4% |
| Spain | 8 | 4% |
| Poland | 7 | 4% |
| Sweden | 7 | 4% |
| Norway | 6 | 3% |
| Canada | 4 | 2% |
| Portugal | 3 | 2% |
| Other countries | 21 | 11% |
| No answer | - | - |

**Source:** authors' elaboration. **Note:** 196 answers analysed. Organisation and firm size are defined based on the number of employees: micro firms have fewer than 10 employees, small firms have between 10 and 49 employees, medium firms have between 50 and 249 employees, and large firms have 250 or more employees. Refer to Table B.1 in Annex B for additional statistics on the target sample composition.

We argue that the experience at CERN contributed to getting the ECRs' occupation and, therefore, the associated wage. The survey responses appear to support this hypothesis (Figure 6). Most of the ECRs stated that their learning experience at CERN helped them find their current job, enter the labour market, and expand their professional network. Additionally, they indicated that their time at CERN facilitated obtaining a position with greater responsibilities more quickly, which had a positive impact on their wages. Only a few cases showed no contribution of CERN training to their career development.

Figure 6. Contribution of research training at CERN to ECRs' professional career.

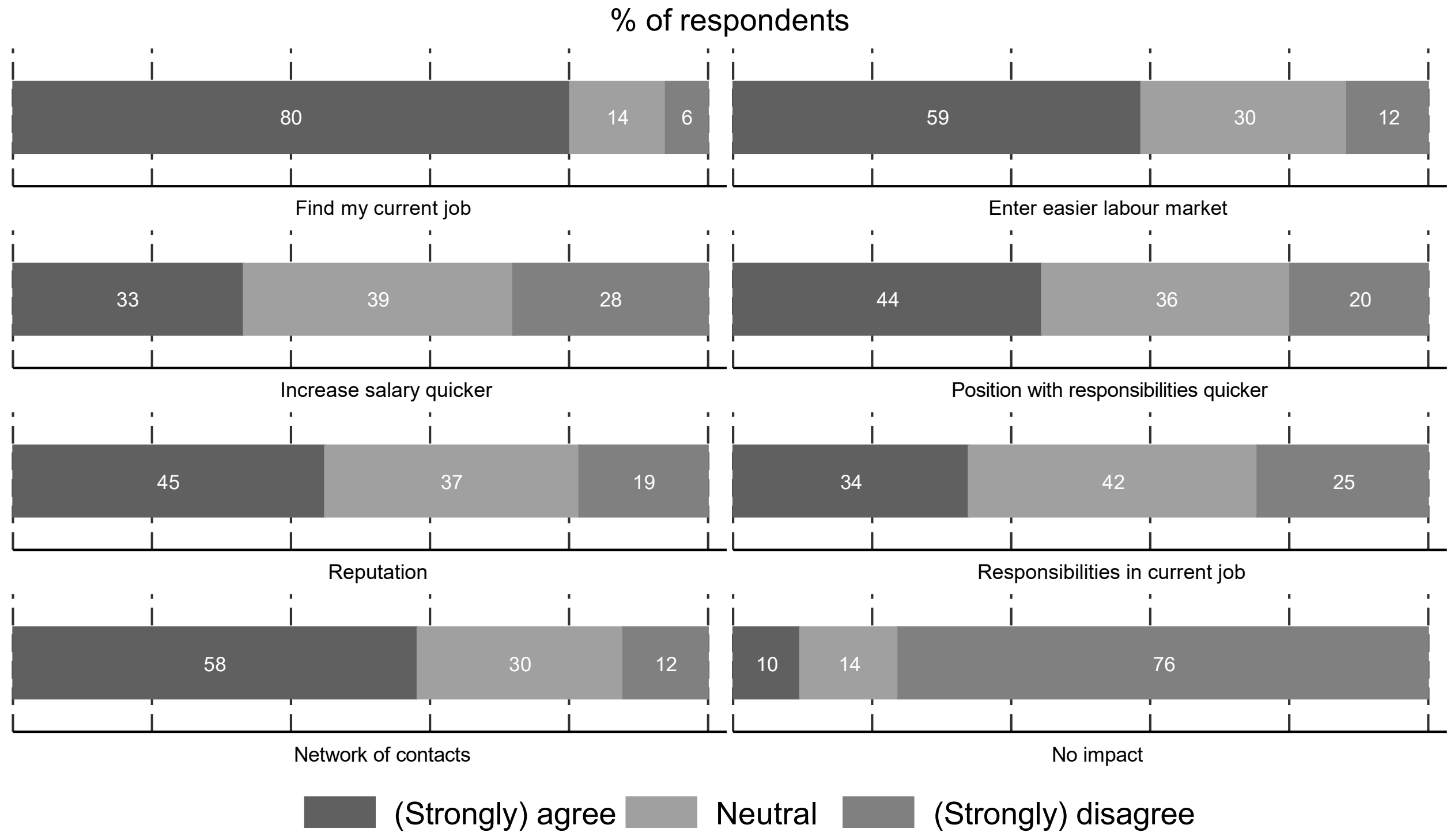

**Source**: authors' elaboration. Note*:* 196 answers analysed. Multiple answers question. Question C.7 of the questionnaire: "For each of the following statements, indicate your level of agreement. My experience at CERN helped me…". Respondents were asked to indicate their answer on a five-point Linkert scale (Strongly disagree, Disagree, Neutral, Agree, Strongly Agree). In the chart, the option "Agree" groups the original options "Strongly Agree" and "Agree" while "Disagree" groups the original options "Strongly disagree" and "Disagree". Refer to Table B.2 in Annex B for further details on the question items.

The observed average yearly gross salary in our sample is EUR 77,401, ranging from a minimum of EUR 6,250 to a maximum of EUR 430,000. Figure B.2 in Annex B shows its log-probability distribution function, while Figure B.3 shows the salary distribution by country of employment. We also asked respondents to indicate the type of contract, [22] whether they received an extra bonus (monetary and in-kind) in addition to salary, which was the case for 51% of ECRs (*n = 101*)[23] and the type of contract with most of the respondents having a full-time contract (*86%; n=168*) (Table B.1, Annex B).

[22] Question B.14 of the questionnaire: "Your labour contract", with options: (i) full-time; (ii) part-time; (iii) temporary contract; (iv) fixed-term contract; (v) seasonal contract; (vi) causal contract and zero hour; (vii) volunteer; (viii) Apprentices.

[23] Question B.12 of the questionnaire: "Do you receive some extra bonus/benefits in addition to your salary?" with options "Yes" and "No", while question B.13: "Your extra bonus is", with options: (i) Monetary; (ii) In-kind (meal voucher, healthcare voucher, etc); (iii) Both monetary and in-kind.

## 4.2 Econometric analysis

We developed a comprehensive set of variables for the econometric analysis by using the survey items (Table 6).

Table 6. Econometric analysis: list of variables

| Survey items | Label (in the model) | Description (as used in the model) |
|---|---|---|
| Current total yearly gross salary | Salary | Hourly gross salary (log) |
| Number of years spent in formal education | Education | Numbers of years of formal education (log) |
| Number of years you have been working since the period at CERN ended | Work experience | Number of years of working experience (log) |
| Other training experiences beyond CERN | Other training | 1 if yes, 0 otherwise |
| Gender | Male | 1 if male, 0 otherwise |
| Age as 2021 | Age | Age as of 2021 (log) |
| Items related to the period at CERN | | |
| Motivation to apply to CERN | Environment, Knowledge, Skill, Job Chance | Set of dummies for each level of the liker scale (from 1 "Not important / Low importance" to 3 "Very Important/ Important ") |
| Start/end – date of the training period at CERN | Duration of the CERN training | Duration in months (log) |
| Skills-related items | No. of hard skills | Number |
| | No. of soft skills | |
| Activity domain when at CERN | Activity domain when at CERN | Set of dummies |
| Salary-related items | | |
| Type of contract | Full-time contract | 1 if full-time, 0 otherwise |
| Recipient of extra bonus in addition to salary | Bonus | 1 if yes, 0 otherwise |
| Context-related items | | |
| Size of the company/organisation | Micro | 1 if micro, 0 otherwise |
| | Small | 1 if small, 0 otherwise |
| | Medium | 1 if medium, 0 otherwise |
| | Large | 1 if large, 0 otherwise |
| Sector of employment | Industry & finance | 1 if yes, 0 otherwise |
| | Academia & research | 1 if yes, 0 otherwise |
| | Public ad, incl. non-profit | 1 if yes, 0 otherwise |
| Country of employment | Country names | For each country: 1 if yes, 0 otherwise |

Source: authors' elaboration

### 4.2.1 Estimation results (H1)

Table 7, Column (1) displays the key coefficients for Eq. (3), where the duration of the CERN learning experience is specified as a linear predictor of salary. Column (2) extends it by including the squared term of that duration, capturing potential non-linearities in the relationship (Eq. 4).[24] Column (3) introduces a broader set of control variables as reported in Eq. (5) and represents our baseline specification. The dependent variable is the ECRs' gross yearly salary measured in euros, while the "active period at CERN" (our main explanatory variable of interest) is measured in months. Both of them are expressed in natural logarithm. We control for a range of additional factors potentially influencing the ECRs' actual salary. These include demographic characteristics and job-related attributes, such as gender, employment sector, organisation size, contract type, the composition of the total reward package, including monetary and in-kind benefits, and the country of employment. Furthermore, the analysis incorporates specific aspects of the training, such as the activity domain in the research project at CERN

Table 7. Estimation results: wage returns to CERN training

| | (1) | (2) | (3) |
|---|---|---|---|
| Variables | Gross yearly salary (log) | Gross yearly salary (log) | Gross yearly salary (log) |
| Active period at CERN (log) | 0.21*** | 0.34 | 0.39* |
| | (0.06) | (0.28) | (0.23) |
| Active period at CERN squared(log) | | -0.02 | -0.04 |
| | | (0.04) | (0.04) |
| Years of education (log) | -0.12 | -0.12 | 0.20* |
| | (0.13) | (0.13) | (0.12) |
| Working experience (log) | 0.15*** | 0.15*** | 0.14*** |
| | (0.05) | (0.05) | (0.04) |
| Other trainings | 0.06 | 0.06 | 0.05 |
| | (0.07) | (0.07) | (0.06) |
| Demographics and job characteristics | | | Yes |
| Activity domain during the active period at CERN | | | Yes |
| Country | Yes | Yes | Yes |
| Constant | 10.93*** | 10.74*** | 9.08*** |
| | (0.40) | (0.61) | (0.59) |
| Observations | 192 | 192 | 189 |
| R-squared | 0.54 | 0.55 | 0.65 |
| F-test | 15.59 | 14.78 | 107.40 |

[24] We also used other different functional forms linking wages to CERN training, including months in units (instead of logarithm) and higher polynomial degrees. Results did not change,

Source: authors' elaboration. Note: The table reports OLS estimations of equations (3), (4), and (5). The dependent variable is the logarithm of the gross annual salary of ECRs. Active period at CERN and Active period at CERN squared refer to the log transformation of the training duration, expressed in months, and its squared term. Years of education represent the total years of formal education in the logarithm. Years of experience denotes the total number of years of work experience in logarithm. Other training is a dummy variable that takes the value of 1 if the ECR attended other training programmes besides the one at CERN. The group "Demographics and Job Characteristics" includes a vector of control variables: gender (1 = male), organisation size (1 = large organisation), sector of employment (1 = academia), type of job contract (1 = full-time contract), and additional bonus (1 = if the reward package includes either monetary or non-monetary bonuses). Activity domain is a set of dummy variables indicating the type of activity performed at CERN. Country is a vector of country dummies. Robust standard errors in parentheses. *** p<0.01, ** p<0.05, * p<0.1.

The duration of the active period at CERN is consistently positive and statistically significant in specifications (1) and (3), suggesting the existence of a wage return to research time spent at CERN, corroborating our hypothesis H1. A 1% increase in the months of training duration is associated with a 0.39% increase in the gross yearly salary. The squared term of it is estimated at -0.04%, and although negative as expected, it is not statistically significant.[25] The cumulative wage returns to the period at CERN values 7% of the annual salary at the average duration of 41 months (3.4 years), with estimation ranging from 2 to 10% (Figure 7, Panel A). Considering the average annual ECRs' salary in our sample of EUR 77,401, the wage gain corresponds to EUR 5,400 per year, ranging from EUR 1,500 to 7,700 mainly materialised within the first ten years of ECRs' working careers.[26] Table 7, Panel B shows the marginal effects of the CERN wage returns for each month of training, which are in line with the empirical evidence of decreasing marginal returns (Leuven, 2004).

The coefficients associated with the other variables have the usual meaning. The number of years of formal education (log) is positive and statistically significant in Column (3), indicating a 0.20% increase in salary for a 1% increase in years of formal education.[27] Working experience (log), modelled as a linear function of wage shows a positive, stable, and significant relationship with salary in all specifications, with coefficients around the value of 0.15. Importantly, the coefficient for other training experiences unrelated to CERN shows no significant relationship with salary across all specifications, highlighting the distinctive impact of the CERN experience on ECRs wages.

Figure 7 – Estimation results: wage returns as function of the active period at CERN

PANEL A: Cumulative wage returns

[25] An F-test of the two terms confirms joint significance: $F(2, 159) = 4.64$; Prob>F=0.0110.

[26] As discussed in Section 4.1, 98% of respondents having a working career of less than 10 years.

[27] The number of years of education is not statistically significant in specifications (1) and (2). This is likely because that 88% of ECRs in our sample has a doctoral or a master degree, and some effects only emerge when the specification is improved with additional controls. On top of that, the high education level of CERN ECRs makes our findings poorly comparable with wage returns to education in the education economic literature, as outlined in earlier sections.

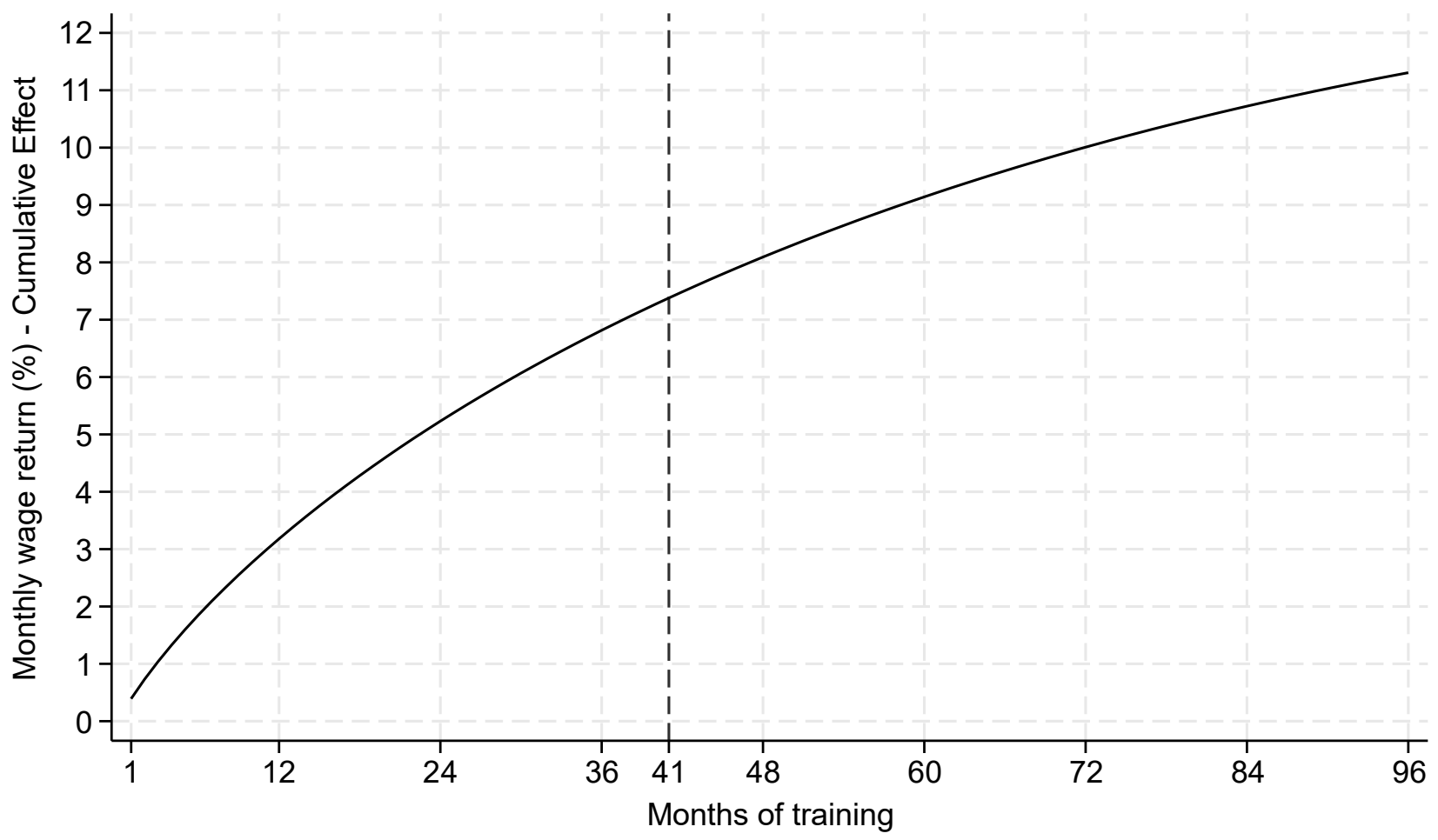

PANEL B: Marginal wage returns

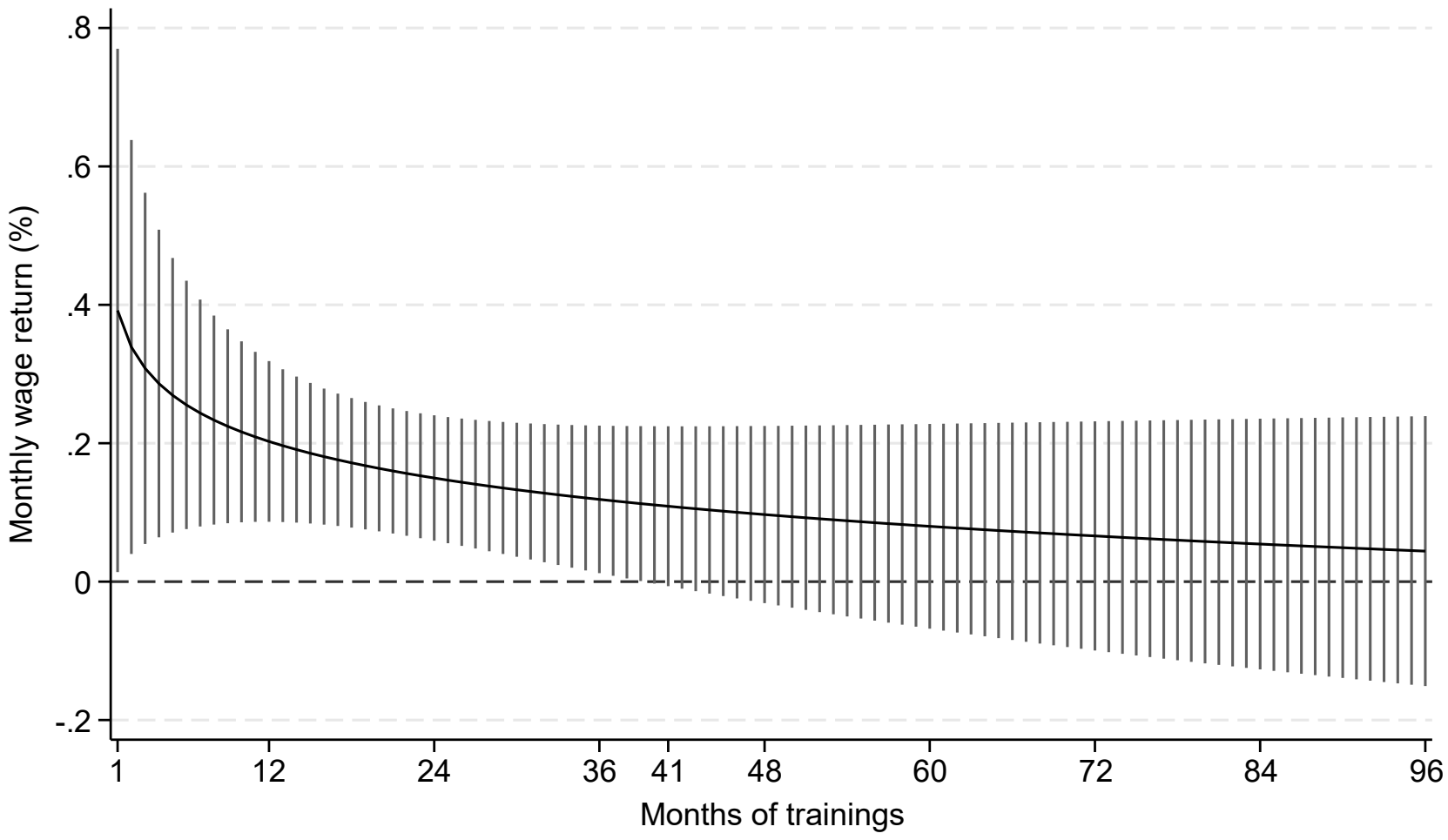

Source: authors' elaboration based on specification 3 in Table 7.

## 4.2.2 Productivity versus reputational effects (H2 and H3)

Hypothesis H2 posits that the skills acquired by ECRs during the period at CERN determine the impact of the training on wages. To test this hypothesis, we estimated the relationships between training duration and the acquisition of hard and soft skills, as outlined in Eq. (7) and (8), respectively. If H2 holds, the coefficients $\gamma_4$ in Eq. (7) and $\delta_4$ in Eq. (8) are expected to be positive and statistically significant.

$$F(Hard\ skills_i) = \gamma_0 + \gamma_1\, EDU_i + \gamma_2\, f(EXP_i) + \gamma_3\, EDUFIELD_i + \gamma_4 f(CERN) + u_i \qquad (7)$$

$$F(Soft\ skills_i) = \delta_0 + \delta_1\, EDU_i + \delta_2\, f(EXP_i) + \delta_3\, EDUFIELD_i + \delta_4 f(CERN) + \eta_i \qquad (8)$$

Results are presented in Table 8 from various models. Column (1) reports the marginal effects from a Logit model estimating the probability of improving at least one skill. Column (2) uses an OLS model, transforming the number of skills acquired into logarithmic form. Columns (3) and (4) employ Poisson and Negative Binomial models, respectively, treating the number of skills as a count variable. In all models, a vector of confounders is included to consider the effect on the skill acquisition of other observables such as gender, educational background (field), years of education, years of work after leaving CERN, and activity domain during training. To address the potential ability bias, we also control whether respondents held a managerial position, which might be considered a proxy of individual abilities that simultaneously influenced the decision to participate in a project at CERN and the likelihood of acquiring or improving skills (e.g., determination, self-esteem).

The logistic regression model indicates that the training duration (and its squared term) is positive and statistically significant, as expected, pointing to a strong association between research time spent at CERN and the probability of acquiring or improving at least one hard skill. The OLS specification corroborates this finding, indicating that a 1% increase in training duration determines a 0.5% increase in the number of hard skills acquired, with diminishing returns of 0.07%. The relationship between the training duration and hard skills is not statistically significant in the Poisson and Negative Binomial specifications. When soft skills are considered, the results closely mirror those obtained for hard skills, with the coefficient being positive and statistically significant across all the specifications except in the logit model (Table 8, Column 5).

Alternatively, one could argue that if a presence at CERN in the frame of a research project enhances certain skills, we would expect the coefficients on skills to lose their statistical significance when both the duration of the presence at CERN and skills are included in the same wage equation. Table 9 reports the estimation of Eq. (6), where the baseline specification is extended to incorporate the skills-related variables measured using different metrics. Column (1) introduces dummy variables indicating whether individuals acquired or improved at least one hard or soft skill. Column (2) includes the total count of acquired skills, while Column (3) uses the logarithm of the number of skills. All specifications control for personal and job-related characteristics as well as the activity domain at CERN. Regardless of the metrics used, the coefficients associated with the skills acquired are either not statistically significant or, in a few cases, show negative signs, reflecting multicollinearity with the duration of the presence at CERN. In other words, the duration of the stay at CERN already accounts for the role of skills in determining ECRs' wages.

Turning to H3, it states that being active in a project at CERN influences ECR's job opportunities through reputation mechanisms, complementing its effects on productivity. A natural identification strategy to distinguish between productivity and screening effect would imply examining the impact of training on individuals' productivity. Unfortunately, no measure of the latter is available. As a result, most empirical studies rely on earnings functions, assuming wages as a proxy for productivity (Psacharopoulos and Patrinos, 2018; Asplund, 2001; Layard and Psacharopoulos, 1974). This creates a challenge because both the screening hypothesis and human capital theory predict that training positively influences earnings. Empirical analyses attempting to discriminate between the two views have developed various identification strategies (see Scalli, 2001 for a review), including estimating wage equations for self-employed workers and employees separately. The idea is that wage returns to education (and training) should be lower for self-employed workers than employees because self-employed people have complete information about their innate ability, and therefore, their earnings are only driven by productivity-related effects (Wolpin 1977).

We included a question to test H3 in our survey (Question B.1, Annex A), but unfortunately, we only received 3 answers (out of 196) from self-employed ECRs, preventing us from statistically disentangle reputational from productivity impacts. In the absence of data, we can claim that the observed wage returns are more likely to stem from productivity effects rather than pure screening mechanisms. The positive relationship between training duration, skill acquisition, and wage increases supports the notion that a presence at CERN in the frame of a project enhances ECRs' productivity, which subsequently influences their salary trajectory, which aligns with the predictions of the human capital accumulation theory.

Table 8. Estimation results: the effect of CERN training on skills

| | Hard Skill | | | | Soft Skills | | | |
|---|---|---|---|---|---|---|---|---|
| | (1) | (2) | (3) | (4) | (5) | (6) | (7) | (8) |
| | (Logit) | (OLS) | (Poisson) | (Negative Binomial) | (Logit) | (OLS) | (Poisson) | (Negative Binomial) |
| Variables | Hard Skills (dummy) | No. Hard Skill (log) | No. Hard Skill | No. Hard Skill | Soft Skill (dummy) | No. Soft Skill (log) | No. Soft Skill | No. Soft Skill |
| Active period at CERN (log) | 69.62** | 0.58** | 0.87 | 0.87 | 2.54 | 0.34* | 0.51** | 0.51** |
| | (129.85) | (0.22) | (0.57) | (0.57) | -4.34 | -0.2 | -0.22 | -0.22 |
| Active period at CERN squared (log) | 0.52** | -0.07** | -0.11 | -0.11 | 0.92 | -0.05 | -0.08** | -0.08** |
| | (0.16) | (0.04) | (0.08) | (0.08) | -0.3 | -0.04 | -0.04 | -0.04 |
| Gender and country of work | Yes | Yes | Yes | Yes | Yes | Yes | Yes | Yes |
| Activity domain at CERN - dummies | Yes | Yes | Yes | Yes | Yes | Yes | Yes | Yes |
| Years of formal education and educational background | Yes | Yes | Yes | Yes | Yes | Yes | Yes | Yes |
| Years of work, managerial role | Yes | Yes | Yes | Yes | Yes | Yes | Yes | Yes |
| $\ln(\alpha)$ | | | | -28.76 | | | | -4.64 |
| | | | | (0.00) | | | | -3.7 |
| Constant | 0.04 | -0.35 | -1.35 | -1.35 | 1.85 | 0.90 | 0.50 | 0.50 |
| | (0.11) | (0.40) | (1.03) | (1.03) | (3.30) | (0.64) | (0.70) | (0.69) |
| Observations | 143 | 192 | 192 | 192 | 112 | 192 | 192 | 192 |
| (Pseudo) R-squared | 0.193 | 0.237 | 0.0511 | 0.0511 | 0.266 | 0.157 | 0.05 | 0.04 |
| F/Wald-Test | 23.98 | 4.688 | 103.6 | 103.6 | 31.33 | 2.22 | 75.07 | 74.02 |
| LL | | | -328.1 | -328.1 | | | -410.1 | -410.0 |
| AIC | | | 718.12 | 718.12 | | | 882.15 | 884.06 |
| BIC | | | 819.10 | 819.10 | | | 983.14 | 988.30 |

Source: Authors' elaboration. Note: The table reports different specifications of equation (7) using Logit, OLS, Poisson, and Negative Binomial models for both hard skills (left panel) and soft skills (right panel). The dependent variables are as follows: Hard and Soft Skills (dummy) takes the value of 1 if the ECR has acquired at least one hard or soft skill; No. of Hard and Soft Skills represents the total number of skills acquired, irrespective of type; and No. of Hard and Soft Skills (log) represents the logarithm of the number of skills plus 1. Active period at CERN and Active period at CERN squared refer to the log transformation of the training duration, expressed in months, and its squared term. Gender and country of work include a vector of country dummies along with a dummy for male respondents. Years of education and educational background include years of formal education, expressed in logarithmic form, and a set of dummies identifying the educational background. Years of work, Managerial role includes years of work experience, expressed in logarithmic form, and a dummy variable that takes the value of 1 if the ECR holds a managerial position. Activity domain is a set of dummy variables indicating the type of activity performed at CERN. Country is a vector of country dummies. Robust standard errors in parentheses. *** p<0.01, ** p<0.05, * p<0.1.

Table 9. Estimation results: the role of skills in determining wage returns to CERN training

| Variables | (1) Gross yearly salary (log) | (2) Gross yearly salary (log) | (3) Gross yearly salary (log) |
|---|---|---|---|
| Active period at CERN (log) | 0.44* | 0.44* | 0.45** |
| | (0.23) | (0.23) | (0.22) |
| Active period at CERN squared (log) | -0.05 | -0.04 | -0.05 |
| | (0.04) | (0.04) | (0.03) |
| Years of formal education (log) | 0.19 | 0.24* | 0.22* |
| | (0.12) | (0.12) | (0.12) |
| Working experience (log) | 0.14*** | 0.13*** | 0.13*** |
| | (0.04) | (0.04) | (0.04) |
| Other training | 0.05 | 0.06 | 0.06 |
| | (0.06) | (0.06) | (0.06) |
| Hard skills (dummy) | -0.11 | | |
| | (0.11) | | |
| Soft skills (dummy) | 0.00 | | |
| | (0.20) | | |
| No. Hard Skills | | -0.05* | |
| | | (0.02) | |
| No. Soft Skills | | 0.00 | |
| | | (0.02) | |
| No. Hard Skills (log) | | | -0.14** |
| | | | (0.07) |
| No. Soft Skills (log) | | | 0.03 |
| | | | (0.08) |
| Demographics and job characteristics | Yes | Yes | Yes |
| Activity domain during period at CERN | Yes | Yes | Yes |
| Country | Yes | Yes | Yes |
| Constant | 9.12*** | 8.94*** | 8.97*** |
| | (0.60) | (0.57) | (0.58) |
| Observations | 189 | 189 | 189 |
| R-squared | 0.65 | 0.66 | 0.657 |
| F-test | 100.97 | 42.44 | 48.67 |

Source: authors' elaboration. Note: The table reports OLS estimations of different specifications of equation (6). The dependent variable is the logarithm of the gross annual salary of ECRs after an active period at CERN. "Active period at CERN' and "Active period at CERN squared refer to the log transformation of the training duration, expressed in months, and its squared term. Hard and Soft Skills (dummy) take the value of 1 if the ECR has acquired at least one hard or soft skill. No. Hard and soft skills refer to the total number of skills acquired, regardless of the specific type. No. of Hard and Soft Skills (log) represents the logarithm of the number of skills plus 1. Years of education represent the total number of years of formal education, expressed in logarithmic form. Years of experience denotes the total number of years of work experience, also expressed in logarithmic form. Other training is a dummy variable that takes the value of 1 if the ECR attended other training programmes in addition to the one at CERN. Demographics and Job Characteristics include a vector of control variables: gender (1 = male), organisation size (1 = large organisation), sector of employment (1 = academia), type of job contract (1 = full-time contract), and additional bonus (1 = if the reward package includes either monetary or non-monetary bonuses). Activity domain is a set of dummy variables indicating the type of activity performed at CERN. Country is a vector of country dummies. Robust standard errors in parentheses. *** p<0.01, ** p<0.05, * p<0.1.

### 4.2.3 Robustness checks

To reinforce the validity of our findings, we conducted a series of robustness checks, including (i) placebo tests; (ii) testing the role of additional training opportunities beyond CERN; (iii) alternative estimation procedures such as structural equation modelling approaches.

Table 10, Column (1) shows an estimation of Eq. (5) based on a placebo sample consisting of researchers that had an active period at CERN but are over 40 years of age in our sample. Economic literature shows that the impact of education (Deming, 2022) and training opportunities (Grosemans et al., 2017) on wages is more relevant for young people at the beginning of their working life than later. As careers proceed, wage returns are less likely to be influenced by training experiences and more likely driven by accumulated working experience. If this holds true, the statistical significance of the coefficient on the active period at CERN is expected to disappear when considering the placebo sample. Our results support the evidence with the coefficient not being statistically different from zero; in contrast, the coefficient for work experience increases by approximately 0.10 compared to the main results presented in Table 7, as expected.

Concerns may also arise that controlling for additional training carried out by our ECRs could bias the estimated effects, given the potential correlation between multiple training experiences or the mediating role of other training opportunities. To address this issue and assess the stability of the coefficient on the active period at CERN, we re-estimated Eq. (5), omitting the variable capturing additional training opportunities besides CERN. Column (2) demonstrates that the coefficients associated with both the linear and the squared functional forms of the active period at CERN remain largely unchanged, confirming the primary role of the research experience at the laboratory in contributing to ECRs' actual wages relative to other training courses.

Table 10 – Robustness checks: placebo tests and alternative specifications

| VARIABLES | (1)<br>ECRs over 40 years old | (2)<br>Omitting "Other training opportunities " |
|---|---|---|
| Active period at CERN (log) | 0.25 | 0.40* |
| | (-0.27) | (0.22) |
| Active period at CERN squared (log) | -0.01 | -0.04 |
| | (-0.04) | (0.04) |
| Years of formal education (log) | 0.23 | 0.18 |
| | (-0.31) | (0.12) |
| Working experience (log) | 0.25*** | 0.13*** |
| | (-0.06) | (0.04) |
| Other training | 0.15 | |
| | (-0.09) | |
| Demographics and job characteristics | Yes | Yes |
| Activity domain during the period at CERN | Yes | Yes |
| Country of Work (Dummy) | Yes | Yes |
| Constant | 9.00*** | 9.17*** |
| | -1.11 | -0.56 |
| Observations | 111 | 190 |
| R-squared | 0.64 | 0.647 |

Source: authors' elaboration. Note: The table reports OLS estimations of different specifications of equation (5). The dependent variable is the logarithm of the gross annual salary of ECRs after an active period at CERN. Column (1) sample includes only researchers over 40. Years of education represent the total number of years of formal education, expressed in logarithmic form. Years of experience denotes the total number of years of work experience, also expressed in logarithmic form. Other training is a dummy variable that takes the value of 1 if the ECR attended other training programmes in addition to the one at CERN. Demographics and Job Characteristics include a vector of control variables: gender (1 = male), organisation size (1 = large organisation), sector of employment (1 = academia), type of job contract (1 = full-time contract), and additional bonus (1 = if the reward package includes either monetary or non-monetary bonuses). Activity domain is a set of dummy variables indicating the type of activity performed at CERN. Country is a vector of country dummies. Robust standard errors are reported in parentheses. *** p<0.01, ** p<0.05, * p<0.1.

One could argue that our set of control variables is still insufficient to fully capture ECRs' unobserved ability, which is likely to jointly influence wages, the opportunity to participate in a project at CERN, and the skills acquired during the training. Controlling for unobservable ability is challenging without longitudinal data (Wooldridge, 2010). However, we address this issue by simultaneously estimating the proposed mechanism in which ECRs' motivations to apply (Figure 1, Section 4.1) influence the skills acquired (both hard and soft), which in turn are linked to the duration of the active period at CERN, ultimately affecting wages.

We tested two alternative structural equation modelling approaches: (i) the Seemingly Unrelated Regression (SUR) method and (ii) the Three-Stage Least Squares (3SLS). The SUR model estimates Eq. (6), (7), and (8) simultaneously as three separated equations but capturing potential interdependencies between the outcomes of interest – wages and acquired skills – by allowing for residual correlations in the error terms $(u_i, \eta_i)$. In contrast, the 3SLS estimator treats the skills in Eq. (6) as endogenous variables correlated with the error term $u_i$. The 3SLS combines the methods of the 2SLS and SUR: in the first stage, the vector of motivations for applying to spend a research training in a projects at CERN – considered as a proxy of unobserved ability – is used as exclusion restrictions, namely, to replace the endogenous variables (the skills) with their predicted values. In the second stage, the SUR method accounts for contemporaneous correlation among the error terms (Wooldridge, 2010).

Results are reported in Table 11. The left panel reports the SUR estimations, confirming our main findings. When using the 3SLS estimator, the statistical significance of the coefficients in the earnings equation improves substantially. The coefficient for the linear term of the active period at CERN increases to 0.73% in the 3SLS model, while the squared term becomes -0.09% and is statistically significant. In the skill equations, the primary mechanism linking the duration of the active period at CERN to wages is confirmed to be the acquisition and improvement of hard skills.

Table 11 – Robustness checks: alternative estimators

| | SUR | | | 3SLS | | |
|---|---|---|---|---|---|---|
| VARIABLES | Gross annual salary (log) | Hard Skills (dummy) | Soft Skills (dummy) | Gross annual salary (log) | Hard Skills (dummy) | Soft Skills (dummy) |
| Active period at CERN (log) | 0.45** | 0.48*** | 0.13 | 0.73*** | 0.27** | -0.02 |
| | (0.20) | (0.15) | (0.17) | (0.20) | (0.13) | (0.13) |
| Active period at CERN squared (log) | -0.05 | -0.07*** | -0.01 | -0.09*** | -0.04* | 0.01 |
| | (0.03) | (0.02) | (0.02) | (0.03) | (0.02) | (0.02) |
| Years of formal education (log) | 0.19* | | | 0.20 | | |
| | (0.11) | | | (0.12) | | |
| Working experience (log) | 0.14*** | 0.02 | -0.02 | 0.16*** | -0.00 | -0.03 |
| | (0.04) | (0.03) | (0.02) | (0.05) | (0.03) | (0.02) |
| Other training | 0.05 | 0.03 | -0.03 | 0.09 | 0.02 | -0.04 |
| | (0.06) | (0.04) | (0.04) | (0.07) | (0.05) | (0.03) |
| Hard Skills (dummy) | -0.12 | | | -0.85** | | |
| | (0.10) | | | (0.39) | | |
| Soft Skills (dummy) | 0.02 | | | 0.72 | | |
| | (0.19) | | | (0.48) | | |
| Demographics and job characteristics | YES | | | YES | | |
| Activity Domain at CERN | YES | | | YES | | |
| Country dummies | YES | YES | YES | YES | YES | YES |
| Education background | | YES | YES | | YES | YES |
| Reason to apply at CERN | | | | | YES | YES |
| Constant | 9.11*** | 0.12 | 0.79*** | 8.63*** | -0.07 | 0.59** |
| | (0.55) | (0.25) | (0.30) | (0.61) | (0.26) | (0.27) |
| Observations | 189 | 189 | 189 | 188 | 188 | 188 |
| R-squared | 0.651 | 0.108 | 0.024 | 0.530 | 0.287 | 0.260 |

Source: authors' elaboration. Note: The table reports OLS estimations of different specifications of equation (5). The dependent variable is the logarithm of the gross annual salary of ECRs after the research training at projects at CERN. "Active period at CERN" is expressed in months. Years of education represent the total number of years of formal education, expressed in logarithmic form. Years of experience denotes the total number of years of work experience, also expressed in logarithmic form. Other training is a dummy variable that takes the value of 1 if the ECR attended other training programmes in addition to the one at CERN. Demographics and Job Characteristics include a vector of control variables: gender (1 = male), organisation size (1 = large organisation), sector of employment (1 = academia), type of job contract (1 = full-time contract), and additional bonus (1 = if the reward package includes either monetary or non-monetary bonuses). Activity domain is a set of dummy variables indicating the type of activity performed at CERN. Country is a vector of country dummies. Robust standard errors are reported in parentheses. *** p<0.01, ** p<0.05, * p<0.1.

# 5. Conclusions

This study underscores the role of RIs as drivers of human capital accumulation, highlighting the wage returns associated with training opportunities for young researchers and the mechanisms underpinning these returns. We designed and developed a research strategy to apply the standards tools of the human capital theory and empirical literature to this recent stream of research. We use the opportunities to participate in a project at CERN as a showcase, but the framework, including the designed questionnaire, can be easily generalised and adapted to other RIs.

In the baseline specification, we found that CERN research training yields a statistically significant 7% increase in ECRs' wages, based on an average training duration of 3.4 years, with an estimated range between 2% and 10%.

The baseline magnitude of wage returns to training identified in this study is lower than earlier estimates provided by Catalano et al. (2021), Camporesi et al. (2017) and Florio et al. (2016), the latter reporting a baseline salary return of 12%. Back-of-the-envelope calculations suggest that the opportunity to participate in a project at CERN yields an average wage increase of EUR 5,400 per year, with a range of EUR 1,500 to 7,700, primarily realised early in the ECRs' professional careers, typically within the first decade of employment. Post-training wage gains are largely driven by hard skills, such as scientific and technical competencies (e.g. software development and data analysis), and soft skills, including communication, networking, and leadership, which are increasingly valued in the labour market (Karaca-Atik et al., 2023; Deming, 2022). Furthermore, we argue that the observed wage return to CERN training is more likely attributable to enhanced productivity stemming from skill acquisition rather than signalling effects associated with a CERN affiliation (Catalano et al., 2021; Bianchin et al., 2019; Anderson et al., 2013).

Our findings have both methodological and policy implications. Methodologically, we improved on earlier studies by focusing exclusively on a sample of young ECRs already in the job market, allowing us to analyse actual salary data rather than expectations. Moreover, our estimation strategy aligns with empirical research on human capital accumulation from training opportunities. By controlling for skill acquisition, training duration, and various confounding factors and addressing ability bias through econometric techniques, we achieved a more robust estimation of wage returns. These methodological advancements provide a replicable framework for future studies on this topic.

When it comes to science and innovation policies, our study supports public investment decisions in world-class research and technological infrastructure as effective tools for enhancing R&I capacity (Draghi, 2024b; OECD, 2019). We demonstrate that investing in RIs increases young researchers' competencies and boosts their productivity. Participating in a project at CERN enhances ECRs' productivity through multiple channels, including providing access to state-of-the-art facilities and resources often beyond the reach of individual countries, opportunities for interdisciplinary and international collaboration with senior researchers, industry and peers, exposure to cutting-edge research topics, and supportive mentorship. Since the 1970s, the economics of education literature has advanced in validating human capital theory, particularly regarding foundational skills like numeracy and literacy. However, the mechanisms for developing "higher-order skills" such as problem-solving, teamwork, and leadership remain less understood (Deming and Silliman, 2024; Deeming, 2022). The example of CERN highlights the importance of activities and services that foster the development of these advanced skills.

Despite its contributions, this study has certain limitations. While the empirical framework follows the human capital theory and its application, the absence of longitudinal data limited our ability to fully address unobservable bias. Although a panel data estimator could not be applied, the main estimation results – validated through extensive robustness checks – consistently demonstrate the effect of training duration on ECRs' wages. Testing alternative functional forms of training duration confirmed the appropriateness of the chosen specification. To better evaluate their impact on ECRs' career trajectories, RIs could implement systems to track individuals who have participated in their projects over time, enabling the development of longitudinal studies and richer data collection. Data gathering should also include information on the ECRs' occupational status, allowing for the disentangling of productivity from reputational mechanisms, which is a relevant topic for world-class research infrastructures.

Second, the lack of a control group prevented us from estimating a "true" salary premium, i.e. the difference between the actual salary experience by the ECRs in our sample and a group of their peers without carrying out practical work in a project at CERN, in the spirit of Angrist and Pischke (2009). ECRs' home institutions, as the primary sources of RIs' ECRs, could serve as a logical starting point for constructing control groups. More advanced econometric counterfactual approaches may reveal lower or higher wage returns from the participation of a project at CERN, depending on the combination of omitted variables bias and measurement errors in the key explanatory variables of our estimates.

# ANNEXES

## ANNEX A. The questionnaire

# VALUE OF TRAINING AT CERN

| SECTION A: Personal information and education background | |
|---|---|
| A.1 Nationality | |
| A.2 Country of residence | |
| A.3 Gender | ☐ Male<br>☐ Female<br>☐ Other |
| A.4 Date of birth | Year / Month / Day |
| A.5 Educational background | ☐ Education<br>☐ Arts and humanities<br>☐ Social science, journalism and information<br>☐ Business, administration and law<br>☐ Natural science, mathematics and statistics<br>☐ Information and Communication Technologies<br>☐ Engineering, manufacturing and construction<br>☐ Agriculture, forestry, fisheries and veterinary<br>☐ Health and welfare<br>☐ Services<br>☐ Other, please specify: ______________________ |
| A.6 Highest level of education | ☐ Secondary professional degree II (without A-level) *[go to A.9]*<br>☐ Secondary diploma II general education (A-levels) *[go to A.9]*<br>☐ Higher professional education *[go to A.9]*<br>☐ University degree below Master *[go to A.9]*<br>☐ Master degree *[go to A.9]*<br>☐ Doctoral degree completed *[go to A.7]*<br>☐ Doctoral degree on-going *[go to A.7]* |
| A.7 Name of university at which you are or were enrolled in a doctoral programme. | ________________________ |
| A.8 Department at the university where you are or were last enrolled | ________________________ |
| A.9 Number of years spent in formal education.<br><br>*[The sum of the number of years spent at primary and secondary school, university,* | _____________________ (number of years) |
| *post-university, including doctoral programme]* | |
| A.10 Have you attended other education /training courses beyond your formal education? | ☐ Yes<br>☐ No |

| SECTION B: Your current job | |
|---|---|
| B.1 Current occupational status | ☐ Looking for a job *[go to section C]*<br>☐ Currently unemployed *[go to section C]*<br>☐ Currently working or being a PhD student at CERN *[go to question B.2]*<br>☐ Self-employed /I have created my own company *[go to question B.2]*<br>☐ Employed *[go to question B.2]*<br>☐ Retired *[go to section C]*<br>☐ Other, please specify: ________________ *[go to question B.2]* |
| B.2 Is your current job your first occupation? | ☐ No<br>☐ Yes |
| B.3 Number of years you have been working since leaving CERN.<br><br>[Insert 0 if you are currently at CERN] | _______________ (number of years) |
| B.4 Country where you currently work. | |
| B.5 Name of your current employer or business<br><br>*[Leave blank if not relevant]* | ______________________ (name) |
| B.6 Name of the department or branch you work in:<br><br>*[Leave blank if not relevant]* | ______________________ (name) |
| B.7 Size of the company/organisation you work at or which you own.<br><br>*[indicate the size of the entire company or group rather than the specific branch at which you work]* | ☐ Micro (less than 10 employees)<br>☐ Small (10 – 49 employees)<br>☐ Medium (50 – 249 employees)<br>☐ Large (250 and more employees) |
| B.8 Type of firm/organisation. | ☐ Public administration<br>☐ Public research (academia, research centres)<br>☐ Private<br>☐ Private non-profit |
| B.9 Sector where your firm/organisation operates.<br><br>*[Multiple choice possible]* | ☐ Agriculture<br>☐ Automotive<br>☐ Biotechnology<br>☐ Chemicals<br>☐ Defence industries<br>☐ Education<br>☐ Electrical and electronic engineering<br>☐ Energy<br>☐ Firearms<br>☐ Food and industry<br>☐ Forestry<br>☐ Fisheries |
| | ☐ Industry, trade and services<br>☐ Construction<br>☐ Wholesale and retail trade<br>☐ Transport<br>☐ Accommodation<br>☐ Food service activities<br>☐ Financial and insurance activities<br>☐ Gambling<br>☐ Healthcare industries<br>☐ ICT<br>☐ Industry<br>☐ Maritime industries<br>☐ Mechanical engineering<br>☐ Medical devices<br>☐ Postal services<br>☐ Pharmaceutical<br>☐ Pressure equipment and gas appliances<br>☐ Raw materials, metals, minerals and forest-based industries<br>☐ Renewable energy<br>☐ Social economy<br>☐ Space<br>☐ Textiles<br>☐ Tourism<br>☐ Toys industries<br>☐ Other, please specify: ______________________ |
| B.10 Average number of hours you work weekly | ________________________ (number of hours) |
| B.11 Your current total yearly gross salary<br><br>*[Specify the local currency]* | ________________________ (Local currency [pulldown]) |
| B.12 Do you receive some extra bonus/benefits in addition to your salary? | ☐ Yes *[ go to question B.13]*<br>☐ No *[ go to question B.14]* |
| B.13 Your extra bonus/benefit is… | ☐ Monetary<br>☐ In-kind (meal voucher, healthcare voucher, etc.)<br>☐ Both monetary and in-kind |
| B.14 Your labour contract | ☐ Full-time<br>☐ Part-time<br>☐ Temporary contract<br>☐ Fixed-term contract<br>☐ Seasonal contract<br>☐ Casual contract and zero hour<br>☐ Volunteer<br>☐ Apprentices |
| B.15 Your current job title | ☐ Software engineer<br>☐ Manufacturing engineer<br>☐ Data Analyst/statistician<br>☐ Manager<br>☐ Consultant<br>☐ Teacher/professor<br>☐ Director<br>☐ Executive<br>☐ Technician<br>☐ Administrative employee<br>☐ Other (please specify) |
| B.16 How many additional income generating activities do you have next to your primary occupation?<br><br>[Insert 0 if not relevant] | ___________________ (indicate the number) |

| SECTION C: Your experience at CERN and impact on your career development | |
|---|---|
| C.1.1 Start date at CERN | Month: __, Year: _ _ _ _ |
| C.1.2 (Expected) End date of the period at CERN | Month: __, Year: _ _ _ _ |
| C.2 Work situation at CERN | ☐ STAFF LD<br>☐ STAFF ID<br>☐ SASS<br>☐ GPRO<br>☐ PJAS<br>☐ CASS<br>☐ VISC<br>☐ COAS<br>☐ USER<br>☐ TRNE<br>☐ ADMI<br>☐ TECH<br>☐ FELLOW<br>☐ DOCT<br>☐ TEMP<br>☐ OTHER, please specify: _____________ |
| C.3 Domain of activity<br>*[Multiple choice possible]* | ☐ Theoretical physics<br>☐ Experimental physics and its applications<br>☐ Particle accelerator physics and its applications<br>☐ Administration<br>☐ Management<br>☐ Health and safety<br>☐ Environment<br>☐ Engineering<br>☐ Informatics<br>☐ Human resources<br>☐ Technology<br>☐ Marketing and communication<br>☐ Legal services<br>☐ Finance |

C.4 How do you rate the importance of the following considerations on your decision to apply for a working experience at CERN?

| | *Not at all important* | *Low importance* | Neutral | Important | *Very important* |
|---|---|---|---|---|---|
| C.4.1 Working in an international environment or gaining experience abroad | ☐ | ☐ | ☐ | ☐ | ☐ |
| C.4.2 Develop new skills | ☐ | ☐ | ☐ | ☐ | ☐ |
| C.4.3 Deepening my knowledge and specific competences of my interest | ☐ | ☐ | ☐ | ☐ | ☐ |
| C.4.4 Increasing my chance to find a job | ☐ | ☐ | ☐ | ☐ | ☐ |
| C.4.5 Other, please specify: | | | | | |

| C.5 Indicate your skills improvement experience for each skill below | | | | | |
|---|---|---|---|---|---|
| Thanks to my experience at CERN, I have improved my: | *Not at all* | Slightly | Somewh at | Much | Very Much |
| C.5.1 Scientific skills | ☐ | ☐ | ☐ | ☐ | ☐ |
| C.5.2 Technical skills | ☐ | ☐ | ☐ | ☐ | ☐ |
| C.5.3 Software development or data analysis skills | ☐ | ☐ | ☐ | ☐ | ☐ |
| C.5.4 Inter-person communication and conflict resolution skills | ☐ | ☐ | ☐ | ☐ | ☐ |
| C.5.5 Problem solving capacity | ☐ | ☐ | ☐ | ☐ | ☐ |
| C.5.6 Language skills | ☐ | ☐ | ☐ | ☐ | ☐ |
| C.5.7 Cultural and social skills | ☐ | ☐ | ☐ | ☐ | ☐ |
| C.5.8 Management including personal self-work management | ☐ | ☐ | ☐ | ☐ | ☐ |
| C.5.9 Teamwork or project leadership | ☐ | ☐ | ☐ | ☐ | ☐ |
| C.5.10 Developing, maintaining and using networks of collaborations | ☐ | ☐ | ☐ | ☐ | ☐ |
| C.5.11 Independent thinking/critical analysis/creativity | ☐ | ☐ | ☐ | ☐ | ☐ |
| C.5.12 Other, please specify: | | | | | |

| | |
|---|---|
| C.6 The number of publications and products which you have authored/co-authored and that have been based on your research at CERN.<br>[Insert 0 if not relevant] | ☐ Articles or scientific papers: _______________ (number)<br>☐ Section/chapter in books: _______________ (number)<br>☐ Intellectual properties (open license such as CC, trademarks, patents, proprietary license such as undisclosed methods and prescriptions, professional secrecy): _______________ (number)<br>☐ Software/applications: _______________ (number)<br>☐ Multimedia products: _______________ (number)<br>☐ Other not scientific media products, please specify: _______________ (number) |

C.7 For each of the following statements, indicate your level of agreement

| My experience at CERN helped (will help) me… | *Strongly disagree* | *Disagree* | *Neutral* | *Agree* | *Strongly agree* |
|---|---|---|---|---|---|
| C.7.1… find my current job | ☐ | ☐ | ☐ | ☐ | ☐ |
| C.7.2 … enter easier the labour market | ☐ | ☐ | ☐ | ☐ | ☐ |
| C.7.3 … increase my salary quicker | ☐ | ☐ | ☐ | ☐ | ☐ |
| C.7.4 … obtain a position with responsibilities quicker | ☐ | ☐ | ☐ | ☐ | ☐ |
| C.7.5 … enter in the labour market thanks to CERN's reputation without any direct impact on my salary | ☐ | ☐ | ☐ | ☐ | ☐ |
| C.7.6 … achieve a position of great responsibility in my current job | ☐ | ☐ | ☐ | ☐ | ☐ |
| C.7.7… be part of a wide network of contacts | ☐ | ☐ | ☐ | ☐ | ☐ |
| C.7.8 None of the above. I think that my experience at CERN have had no impact on my current job/salary/position. | ☐ | ☐ | ☐ | ☐ | ☐ |

| SECTION D: Follow-up | |
|---|---|
| Would you like to participate in the next round of this survey foreseen within one year? | ☐ Yes *[please, provide us with your email]*<br>☐ No *[end of the questionnaire]* |
| Your email | |

THANK YOU FOR YOUR PARTICIPATION IN THE SURVEY

You can add comments in the space below:

## ANNEX B. Additional statistics

Table B.1 – Descriptive statistics for the target sample of ECRs

| Variables | Obs. | Mean | Std. dev. | Min | Max |
|---|---|---|---|---|---|
| Demographics | | | | | |
| Age as 2021 | 196 | 34 | 3.72 | 24 | 40 |
| Male | 196 | 0.85 | - | 0 | 1 |
| Years of education | 196 | 19.82 | 3.67 | 8.65 | 37 |
| Doctoral degree | 196 | 0.57 | - | 0 | 1 |
| Doctoral degree ongoing (at the time of the survey) | 196 | 0.03 | - | 0 | 1 |
| Higher professional education | 196 | 0.03 | - | 0 | 1 |
| Master's degree | 196 | 0.32 | - | 0 | 1 |
| University degree (below master) | 196 | 0.05 | - | 0 | 1 |
| Secondary diploma | 196 | 0.01 | - | 0 | 1 |
| Other trainings | 194 | 0.56 | - | 0 | 1 |
| Work-related information | | | | | |
| Years of work experience | 196 | 4.10 | 2.57 | 0 | 11 |
| Company size: Micro | 193 | 0.05 | - | 0 | 1 |
| Company size: Small | 193 | 0.6 | - | 0 | 1 |
| Company size: Medium | 193 | 0.13 | - | 0 | 1 |
| Company size: Large | 193 | 0.77 | - | 0 | 1 |
| Sector of employment: Industry & finance | 196 | 0.51 | - | 0 | 1 |
| Sector of employment: Academia & research | 196 | 0.41 | - | 0 | 1 |
| Sector of employment: Public ad, incl. non-profit | 196 | 0.09 | - | 0 | 1 |
| Type of contract: full time | 196 | 0.86 | - | 0 | 1 |
| Gross yearly salary (EUR) | 196 | 77,796 | | 6,250 | 430,000 |
| Bonus on top of the salary | 196 | 0.51 | - | 0 | 1 |
| Country of Employment | | | | | |
| Switzerland | 196 | 0.23 | - | 0 | 1 |
| Germany | 196 | 0.11 | - | 0 | 1 |
| France | 196 | 0.08 | - | 0 | 1 |
| United States | 196 | 0.08 | - | 0 | 1 |
| United Kingdom | 196 | 0.07 | - | 0 | 1 |
| Italy | 196 | 0.06 | - | 0 | 1 |
| Netherlands | 196 | 0.05 | - | 0 | 1 |
| Austria | 196 | 0.04 | - | 0 | 1 |
| Spain | 196 | 0.04 | - | 0 | 1 |
| Sweden | 196 | 0.04 | - | 0 | 1 |
| Poland | 196 | 0.04 | - | 0 | 1 |
| Norway | 196 | 0.03 | - | 0 | 1 |
| Canada | 196 | 0.02 | - | 0 | 1 |
| Portugal | 196 | 0.02 | - | 0 | 1 |
| Other countries | 196 | 0.11 | - | 0 | 1 |

Source: authors' elaboration. Note: Rounded figures.

Figure B.1 – Work experience distribution (*N* = *195*)

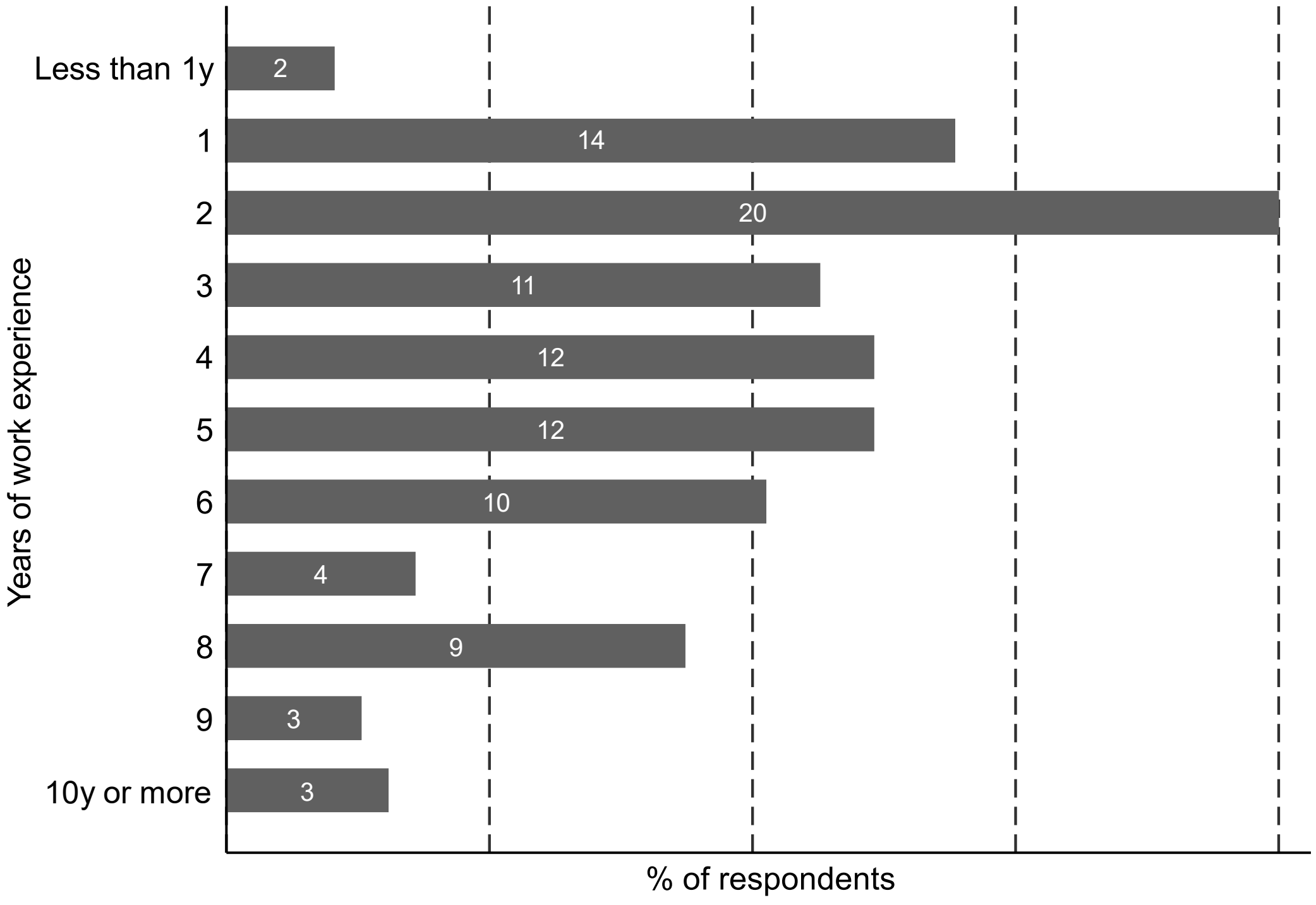

Source: authors' elaboration. Note: Round figures.

Figure B.2– Yearly salary distribution (log)

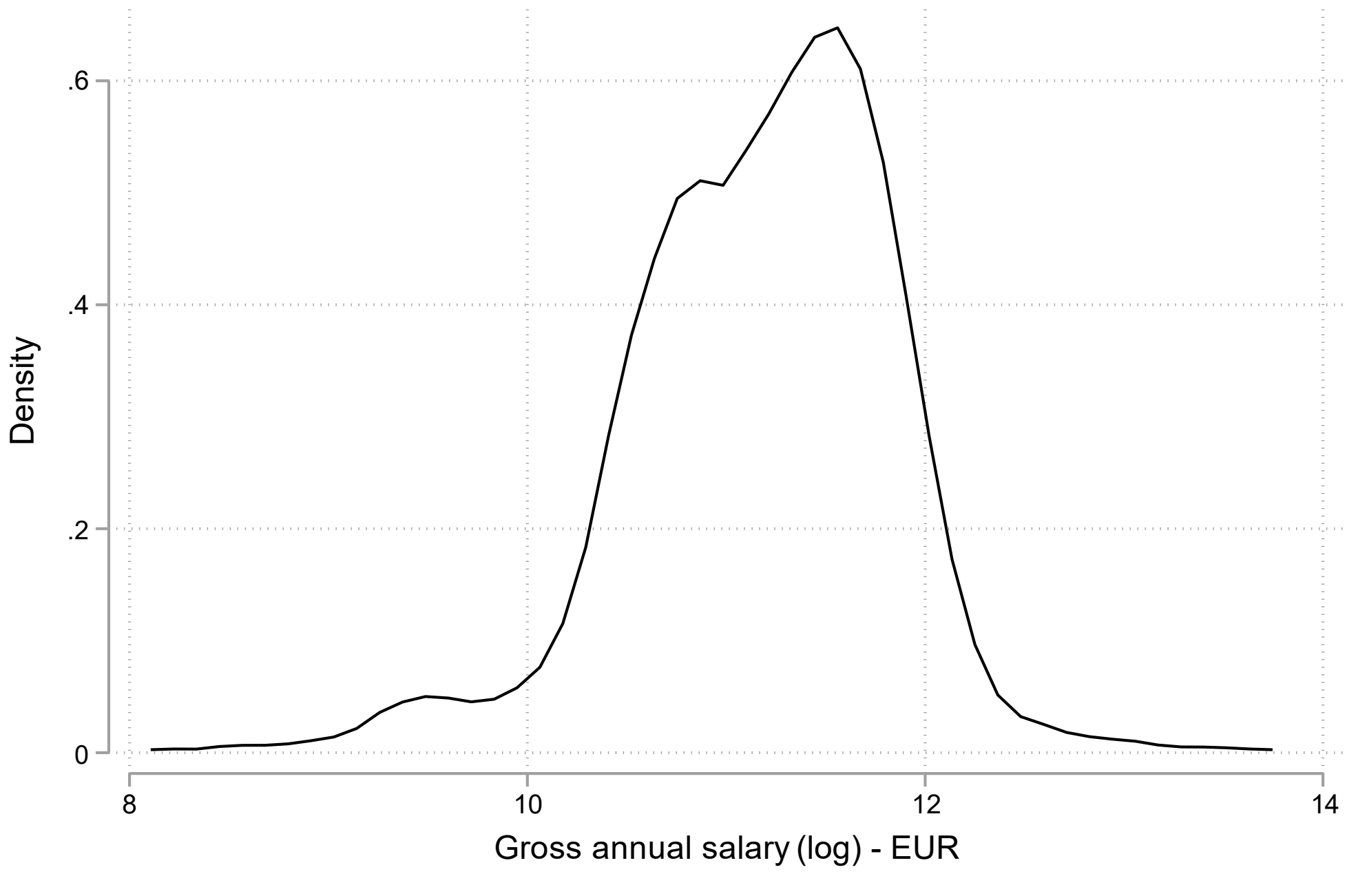

Source: authors' elaboration

Figure B.3 – Average gross yearly salary by country, EUR (*N = 196*)

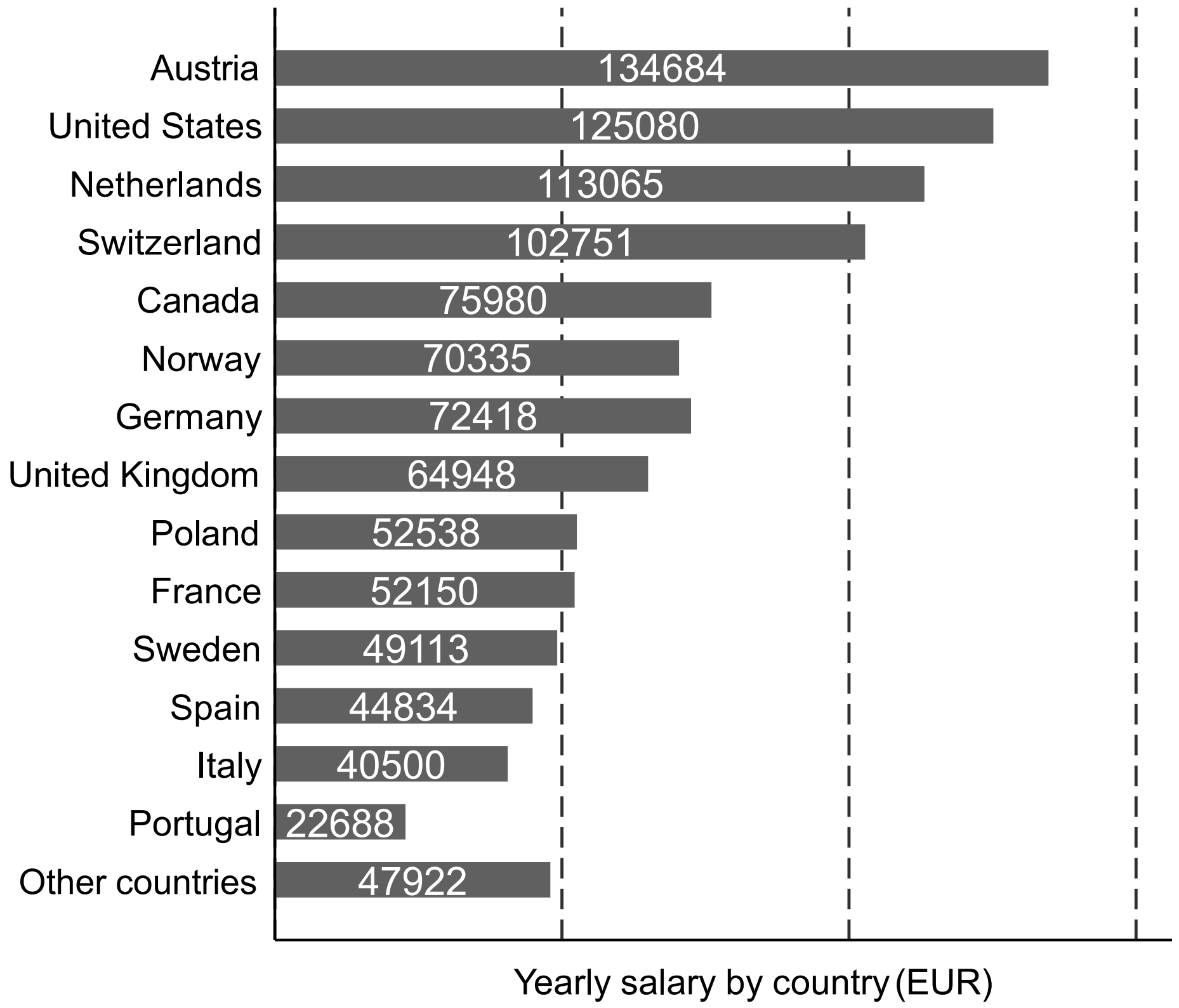

Source: authors' elaboration. Note: Round figures.

Table B.2 – Descriptive statistics on training information

| Variables | Obs. | Mean | Std. dev. | Min | Max |
|---|---|---|---|---|---|
| Duration | | | | | |
| Training duration (years) | 196 | 3.43 | 1.70 | 0.08 | 8 |
| Training duration (months) | 196 | 41.18 | 20.38 | 1 | 99 |
| Domain of activity | | | | | |
| Experimental Physics | 195 | 0.32 | - | 0 | 1 |
| Engineering | 195 | 0.28 | - | 0 | 1 |
| Particle accelerator physics and techno | 195 | 0.14 | - | 0 | 1 |
| Informatics | 195 | 0.13 | - | 0 | 1 |
| Admin, Finance, HR, HSE, Manag. and marketing | 195 | 0.07 | - | 0 | 1 |
| Theoretical physics | 195 | 0.06 | - | 0 | 1 |
| Reasons to apply (Likert scale from 1 "Not important / Low importance" to 3 "Very Important/ Important ") | | | | | |
| Working in an international environment or gaining experience abroad | 195 | 2.92 | 0.32 | 1 | 3 |
| Develop new skills | 194 | 2.89 | 0.36 | 1 | 3 |
| Deepening my knowledge and specific competences of my interest | 195 | 2.83 | 0.48 | 1 | 3 |
| Increasing my chance to find a job | 195 | 2.49 | 0.70 | 1 | 3 |
| Skill acquired (Likert scale from 1 "(Very) Much to 3 "Slightly / Not at all") | | | | | |
| Scientific skills | 195 | 1.39 | 0.68 | 1 | 3 |
| Technical skills | 196 | 1.74 | 0.82 | 1 | 3 |
| Problem solving | 196 | 1.36 | 0.65 | 1 | 3 |
| Communication | 196 | 1.53 | 0.73 | 1 | 3 |
| Language skills | 196 | 1.48 | 0.72 | 1 | 3 |
| Cultural and social skills | 196 | 1.97 | 0.79 | 1 | 3 |
| (Self-work) Management skills | 196 | 1.70 | 0.80 | 1 | 3 |
| Teamwork or project leadership | 196 | 1.73 | 0.79 | 1 | 3 |
| Critical analysis | 196 | 1.49 | 0.73 | 1 | 3 |
| CERN contribution to career (Likert scale from 1 "Strongly agree / agree" to 3 "Strongly disagree / disagree") | | | | | |
| Find my current job | 196 | 1.26 | 0.56 | 1 | 3 |
| Enter easier the labour market | 196 | 1.53 | 0.70 | 1 | 3 |
| Increase my salary quicker | 196 | 1.95 | 0.78 | 1 | 3 |
| Obtain a position with responsibilities quicker | 196 | 1.76 | 0.77 | 1 | 3 |
| Enter in the labour market thanks to CERN's reputation without any direct impact on my salary | 194 | 1.74 | 0.75 | 1 | 3 |
| Achieve a position of great responsibility in my current job | 195 | 1.91 | 0.76 | 1 | 3 |
| Be part of a wide network of contacts | 196 | 1.54 | 0.70 | 1 | 3 |
| I think that my experience at CERN have had no impact on my current job/salary/position. | 177 | 2.67 | 0.65 | 1 | 3 |

Source: authors' elaboration. Note: Round figures.

# ANNEX C- CERN LHC ECRs' affiliation institutions

**Table C.1 – CERN LHC ECRs' affiliation institutions** (*n = 196).*

| Organisation | No. | Percent | Cum. |
|---|---|---|---|
| EPFL (École polytechnique fédérale de Lausanne) | 7 | 5.93 | 5.93 |
| Technical University of Vienna | 5 | 4.24 | 10.17 |
| University of Bonn | 4 | 3.39 | 13.56 |
| Université Paris-Saclay | 3 | 2.54 | 16.1 |
| Warsaw University of Technology | 3 | 2.54 | 18.64 |
| California Institute of Technology | 2 | 1.69 | 20.33 |
| ETH Zurich | 2 | 1.69 | 22.02 |
| KU Leuven | 2 | 1.69 | 23.71 |
| Maynooth University | 2 | 1.69 | 25.4 |
| Politecnico di Milano | 2 | 1.69 | 27.09 |
| University of Manchester | 2 | 1.69 | 28.78 |
| University of Pavia | 2 | 1.69 | 30.47 |
| University of Perugia | 2 | 1.69 | 32.16 |
| Vrije Universiteit Brussel | 2 | 1.69 | 33.85 |
| Carnegie Mellon University | 1 | 0.85 | 34.7 |
| Centro de investigación y estudios av.. | 1 | 0.85 | 35.55 |
| Columbia University | 1 | 0.85 | 36.4 |
| Comenius University | 1 | 0.85 | 37.25 |
| ESIA | 1 | 0.85 | 38.1 |
| Federico II University | 1 | 0.85 | 38.95 |
| Ghent University | 1 | 0.85 | 39.8 |
| Goethe University Frankfurt | 1 | 0.85 | 40.65 |
| Graz University of Technology | 1 | 0.85 | 41.5 |
| Heidelberg University | 1 | 0.85 | 42.35 |
| Humboldt University of Berlin | 1 | 0.85 | 43.2 |
| Imperial College London | 1 | 0.85 | 44.05 |
| Indian Institute of Science | 1 | 0.85 | 44.9 |
| Institute of Nuclear Physics Polish A.. | 1 | 0.85 | 45.75 |
| Jagiellonian University | 1 | 0.85 | 46.6 |
| Johannes Gutenberg University Mainz | 1 | 0.85 | 47.45 |
| Karlsruhe Institute of Technology | 1 | 0.85 | 48.3 |
| Lancaster University | 1 | 0.85 | 49.15 |
| Luleå University of Technology | 1 | 0.85 | 50 |
| Northern Illinois Universtiy | 1 | 0.85 | 50.85 |
| Oxford University | 1 | 0.85 | 51.7 |
| Polytechnic University of Madrid | 1 | 0.85 | 52.55 |
| QUT | 1 | 0.85 | 53.4 |
| RWTH Aachen | 1 | 0.85 | 54.25 |
| Rutgers University | 1 | 0.85 | 55.1 |
| Sapienza University of Rome | 1 | 0.85 | 55.95 |
| Scuola Normale Superiore di Pisa | 1 | 0.85 | 56.8 |
| Stockholm University | 1 | 0.85 | 57.65 |
| TU Eindhoven | 1 | 0.85 | 58.5 |
| TU Munich | 1 | 0.85 | 59.35 |
| TU-Darmstadt | 1 | 0.85 | 60.2 |
| Technical University of Denmark | 1 | 0.85 | 61.05 |
| Torino | 1 | 0.85 | 61.9 |
| UC Berkeley | 1 | 0.85 | 62.75 |
| Universidad de Los Andes | 1 | 0.85 | 63.6 |
| Universidad de Los Andes | 1 | 0.85 | 64.45 |
| Universidad de Zaragoza | 1 | 0.85 | 65.3 |

| Organisation | No. | Percent | Cum. |
|---|---|---|---|
| Universita di Roma "Tor Vergata" | 1 | 0.85 | 66.15 |
| Universitat Jaume I | 1 | 0.85 | 67 |
| Universitatea Transilvania din Brasov | 1 | 0.85 | 67.85 |
| Universite Libre de Bruxelles | 1 | 0.85 | 68.7 |
| University of Manchester | 1 | 0.85 | 69.55 |
| University of Amsterdam | 1 | 0.85 | 70.4 |
| University of Barcelona | 1 | 0.85 | 71.25 |
| University of Bergen | 1 | 0.85 | 72.1 |
| University of Bologna | 1 | 0.85 | 72.95 |
| University of California | 1 | 0.85 | 73.8 |
| University of Cambridge | 1 | 0.85 | 74.65 |
| University of Edinburgh | 1 | 0.85 | 75.5 |
| University of Glasgow | 1 | 0.85 | 76.35 |
| University of Heidelberg | 1 | 0.85 | 77.2 |
| University of Innsbruck | 1 | 0.85 | 78.05 |
| University of Lisbon | 1 | 0.85 | 78.9 |
| University of Liverpool | 1 | 0.85 | 79.75 |
| University of Macedonia | 1 | 0.85 | 80.6 |
| University of Michigan | 1 | 0.85 | 81.45 |
| University of Notre Dame | 1 | 0.85 | 82.3 |
| University of Oregon | 1 | 0.85 | 83.15 |
| University of Oslo | 1 | 0.85 | 84 |
| University of Oviedo | 1 | 0.85 | 84.85 |
| University of Potsdam | 1 | 0.85 | 85.7 |
| University of Rome "Tor Vergata" | 1 | 0.85 | 86.55 |
| University of Sheffield | 1 | 0.85 | 87.4 |
| University of Silesia | 1 | 0.85 | 88.25 |
| University of Stanford | 1 | 0.85 | 89.1 |
| University of Sussex | 1 | 0.85 | 89.95 |
| University of Tokyo | 1 | 0.85 | 90.8 |
| University of Toulouse | 1 | 0.85 | 91.65 |
| University of Warwick | 1 | 0.85 | 92.5 |
| University of Washington | 1 | 0.85 | 93.35 |
| University of Wisconsin-Madison | 1 | 0.85 | 94.2 |
| University of York | 1 | 0.85 | 95.05 |
| University of Zagreb | 1 | 0.85 | 95.9 |
| University vienna | 1 | 0.85 | 96.75 |
| Univesitat de Valencia | 1 | 0.85 | 97.6 |
| Vrije Universiteit Amsterdam | 1 | 0.85 | 98.45 |
| École Centrale de Lyon | 1 | 0.85 | 99.3 |

**Source:** authors' elaboration.